\documentclass[aps,prb,citeautoscript,twocolumn,showpacs,superscriptaddress]{revtex4-1} 
\usepackage{graphicx}
\usepackage{amsmath}
\usepackage{color}
\usepackage[retainorgcmds]{IEEEtrantools}
\begin{document}

\title{Candidate Quantum Spin Ice in Pyrochlore Pr$_2$Hf$_2$O$_7$}
\author{Romain Sibille}
\email[]{romain.sibille@psi.ch}
\affiliation{Laboratory for Scientific Developments and Novel Materials, Paul Scherrer Institut, 5232 Villigen, Switzerland}
\author{Elsa Lhotel}
\email[]{elsa.lhotel@neel.cnrs.fr}
\affiliation{Institut N\'eel, CNRS and Universit\'e Joseph Fourier, BP 166, 38042 Grenoble Cedex 9, France}
\author{Monica Ciomaga Hatnean}
\affiliation{Physics Department, University of Warwick, Coventry, CV4 7AL, UK}
\author{Geetha Balakrishnan}
\affiliation{Physics Department, University of Warwick, Coventry, CV4 7AL, UK}
\author{Bj\"{o}rn F\r{a}k}
\affiliation{Institut Laue-Langevin, CS 20156, F-38042 Grenoble Cedex 9, France}
\author{Nicolas Gauthier}
\affiliation{Laboratory for Scientific Developments and Novel Materials, Paul Scherrer Institut, 5232 Villigen, Switzerland}
\author{Tom Fennell}
\email[]{tom.fennell@psi.ch}
\affiliation{Laboratory for Neutron Scattering and Imaging, Paul Scherrer Institut, 5232 Villigen, Switzerland}
\author{Michel Kenzelmann}
\affiliation{Laboratory for Scientific Developments and Novel Materials, Paul Scherrer Institut, 5232 Villigen, Switzerland}

\begin{abstract}
We report the low temperature magnetic properties of the pyrochlore Pr$_2$Hf$_2$O$_7$. Polycrystalline and single-crystal samples are investigated using time-of-flight neutron spectroscopy and macroscopic measurements, respectively. The crystal-field splitting produces a non-Kramers doublet ground state for Pr$^{3+}$, with Ising-like anisotropy. Below 0.5 K ferromagnetic correlations develop, which suggests that the system enters a spin ice-like state associated with the metamagnetic behavior observed at $\mu_0H_c\sim2.4$~T. In this regime, the development of a discrete inelastic excitation in the neutron spectra indicates the appearance of spin dynamics which are likely related to cooperative quantum fluctuations.
\end{abstract}

\maketitle
\section{Introduction}
Frustration is an important paradigm to stabilize spin liquids\cite{Balents:2010jx,Normand:2009wk}. 
Of special interest are quantum spin liquids (QSLs), which arise from long range entanglement in the ground state wavefunction\cite{Isakov:2011fz,Wen:2002cy}, and which host fractionalized quasiparticle excitations\cite{Gingras:2014ip,Moessner:2010ev}.
In three dimensions, rare-earth pyrochlores\cite{Gardner:2010fu} have proven themselves as a wonderful landscape for discovering and studying exotic spin liquids. The classical spin-ice state\cite{Bramwell2001,Bramwell2001b} arises when ferromagnetic interactions couple the first-neighbor Ising-like moments occupying the  lattice of corner-sharing tetrahedra.  A local constraint - the ice rule - creates a manifold of degenerate ground states that can be regarded as a vacuum in which quasiparticle spin excitations - emergent magnetic monopoles - are created and annihilated\cite{Castelnovo:2008hb}. The nature of the spin correlations generated by the ice rule makes the system a Coulomb phase\cite{Fennell2009,Henley:2010vo} with emergent magnetostatics\cite{Castelnovo:2008hb,Castelnovo:2012kk}.  
An important open question concerns the nature of the spin liquid state when quantum fluctuations amongst the states of the manifold are allowed. By introducing transverse terms in the effective spin-1/2 Hamiltonian on the pyrochlore lattice \cite{Curnoe:2008gy,Ross:2011tv}, the quantum spin ice (QSI) state, a special QSL whose emergent properties are those of a $U(1)$-gauge theory\cite{Hermele:2004gg}, can be stabilized~\cite{Savary:2012cq, Benton:2012ep,Gingras:2014ip}. It exhibits emergent quantum electrodynamics with excitations playing the role of monopoles and photons.

Different materials are seen as candidates for QSL phases on the pyrochlore lattice. They are based on Yb$^{3+}$\cite{Ross:2011tv,Chang:2012el,Applegate2012}, though this case is under debate \cite{Robert2015,Jaubert2015}, Pr$^{3+}$\cite{Zhou:2008cz,Matsuhira2009,Onoda2010,Lee2012,Kimura:2013gj}, or Ce$^{3+}$\cite{Sibille2015}, because for small magnetic moments the transverse terms are not overwhelmed by the dipolar interaction which leads to a classical spin ice.
Two Pr$^{3+}$ pyrochlores were studied so far. Pr$_2$Sn$_2$O$_7$ shows short-range correlations and appears dynamic for inelastic neutron scattering (so far down to 0.2~K) \cite{Zhou:2008cz}. In Pr$_2$Zr$_2$O$_7$ the wave-vector dependence of the elastic part of the magnetic neutron scattering measured at 0.05 K displays features suggesting an ice rule is operative, but these features broaden in the inelastic channels, and it is proposed to be due to the existence of quantum fluctuations out of a QSI state\cite{Kimura:2013gj}. However, the true nature of its ground state is still not fully understood.
Here we report on another chemical variant, Pr$_2$Hf$_2$O$_7$, which is also of interest as a model material to study the QSI since this ground state should be sensitive to the nonmagnetic cation which influences the superexchange paths and the crystal field.
We determine the crystal field parameters from inelastic neutron scattering and we show that the single-ion state of Pr$^{3+}$ in Pr$_2$Hf$_2$O$_7$ possesses the ingredients for the existence of significant residual transverse spin fluctuations. Below $0.5$~K the system likely enters a state related to a QSI, which manifests itself through the appearance of metamagnetism. Inelastic neutron scattering measurements confirm the existence of a dynamical correlated state which we believe arising from quantum fluctuations.

\section{Methods}

The magnetic properties of rare-earth pyrochlore hafnates remain little studied\cite{Craig2009,Anand2015}, at least partially due to their high melting temperatures which makes the preparation of single-crystals less accessible than in the case of titanates. We have recently succeeded in growing single-crystals of Pr$_2$Hf$_2$O$_7$ using the optical floating zone technique\cite{Ciomaga2015}, with xenon lamps in order to attain the temperature required to melt the material\cite{Shevchenko1984}. The powder used as starting material for the growth and as a sample for some of the characterizations reported here was prepared by standard solid-state chemistry techniques, using pre-annealed Pr$_6$O$_{11}$ and HfO$_2$ as starting reagents and applying previously reported annealing conditions\cite{Karthik2012}. A nonmagnetic reference sample, La$_2$Hf$_2$O$_7$, was prepared via the same route, starting from La$_2$O$_3$. Both powder samples were characterized by x-ray diffraction at room temperature and found to be single phase with the Pr$_2$Hf$_2$O$_7$ lattice parameter refined to $10.6842\pm0.0002$ \AA, in agreement with other reports which have established that it crystallizes in the pyrochlore structure\cite{Karthik2012,Blanchard2013}. A parallelepiped sample of mass $m=79.5$ mg was cut from a single crystal of several grams, oriented using a Laue camera, and used for macroscopic measurements.

Magnetization ($M$) data were measured for a polycrystalline sample in the temperature ($T$) range from 1.8 to 370 K in an applied magnetic field ($\mu_0H$) of 0.1 T using a Quantum Design MPMS-XL superconducting quantum interference device (SQUID) magnetometer. Additional magnetization and $ac$-susceptibility measurements were made on the single crystalline sample as a function of temperature and field, from $T$ = 0.07 to 4 K and from $\mu_0H$ = 0 to 8 T, using SQUID magnetometers equipped with a miniature dilution refrigerator developed at the Institut N\'eel-CNRS Grenoble\cite{Paulsen01}. Magnetization data were corrected for demagnetization effects\cite{Aharoni1998}. The heat capacity ($C_p$) of the same sample was measured down to 0.35 K using a Quantum Design physical properties measurement system (PPMS). 

Neutron scattering measurements with thermal and cold neutrons were made on powder samples using the time-of-flight spectrometers IN4 (Institut Laue Langevin, Grenoble, France) and FOCUS (SINQ, Paul Scherrer Institut, Villigen, Switzerland), respectively. On IN4 both samples, Pr$_2$Hf$_2$O$_7$ and La$_2$Hf$_2$O$_7$, were enclosed in flat plate holders made of aluminum and measured using incident energies $E_i$ = 65.7, 116 and 150 meV at $T$ = 1.6 K. An additional spectrum was recorded for Pr$_2$Hf$_2$O$_7$ using $E_i$ = 13 meV in order to verify the absence of crystal field levels in the energy range below 10 meV. On FOCUS we made two experiments; one using a standard `Orange' cryostat and the sample in a flat plate aluminum holder; and another using a dilution refrigerator with the sample in a cylindrical copper can with annular geometry. Spectra were recorded with $E_i$ = 2.47 meV, at several temperatures in the range between $T$ = 1.5 and 100 K for the first experiment and between $T$ = 0.05 and 1.5 K for the second experiment.

\section{Results and discussion}

\subsection{Crystal-field states}

\begin{figure}[ht]
\includegraphics[width=\linewidth]{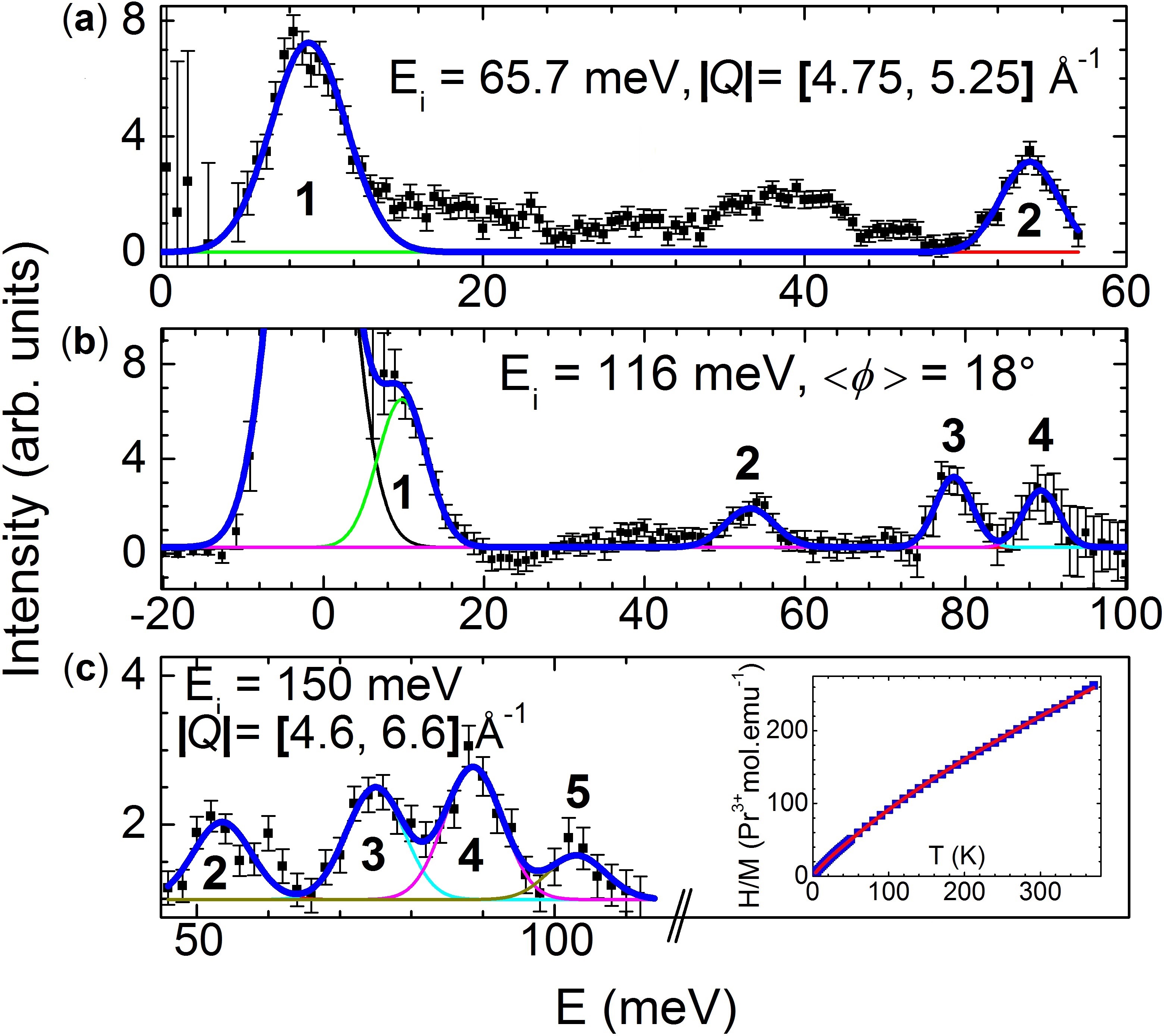}
\centering
\caption{Magnetic neutron scattering intensity as a function of energy transfer at 1.6 K, measured at different incident energies $E_i$. 
In \textbf{a} and \textbf{c} data measured for La$_2$Hf$_2$O$_7$ was subtracted from that of Pr$_2$Hf$_2$O$_7$, and the resulting scattering integrated over a small range of momentum transfer $Q$. In \textbf{b}, high-angle data from Pr$_2$Hf$_2$O$_7$ was subtracted from low-angle data, using a scale factor determined from La$_2$Hf$_2$O$_7$ data~\cite{Princep2013}. Spectra were fitted with a series of Gaussian peaks for the crystal-field transitions labelled \textbf{1}-\textbf{5}. Unlabeled signals are attributed to residual phonon scattering from Pr$_2$Hf$_2$O$_7$ or hydrogen-containing contaminants formed through exposure to air. Note that in \textbf{b} the $|Q|$ range is not constant as a function of energy transfer and integrated intensities have to be extrapolated to $Q = 0$ by correction for the magnetic form factor. The inset in \textbf{c} shows the powder-averaged inverse magnetic susceptibility calculated from the fitted crystal-field parameters (red), and our powder measurement of $H/M$ at $\mu_0H=0.1$~T (blue).}
\label{Fig.1}
\end{figure}

\begin{table}
\begin{ruledtabular}
\begin{tabular}{lccccc}
Level & Degeneracy & $E_{obs}$         & $E_{calc}$      & $I_{obs}$       & $I_{calc}$    \\ \hline
$\Gamma_3^+$     & 2          & 0.0       & 0.0    & -       & -    \\
$\Gamma_1^+$     & 1          & 9.2(3)    & 9.29   & 1.00    & 1.00 \\
$\Gamma_3^+$     & 2          & 53.5(5)   & 53.41  & 0.33(8) & 0.35 \\
$\Gamma_1^+$     & 1          & 78.6(3)   & 78.59  & 0.40(6) & 0.32 \\
$\Gamma_3^+$     & 2          & 89.4(4)   & 89.44  & 0.37(7) & 0.60 \\
$\Gamma_2^+$     & 1          & 103.6(10) & 103.84 & 0.26(8) & 0.26 \\
\end{tabular}
\end{ruledtabular}
\caption{\label{TableI}Observed and calculated crystal-field transition energies and intensities of Pr$_2$Hf$_2$O$_7$. The best crystal-field parameters used for the calculations are $B_0^2 = 34.2$, $B_0^4 = 402.8$, $B_3^4 = 170.8$, $B_0^6 = 147.1$, $B_3^6 = -106.8$, $B_6^6 = 155.4$ meV.}
\end{table}

We first report on the crystal field splitting of Pr$^{3+}$ in Pr$_2$Hf$_2$O$_7$.  By comparing data from Pr$_2$Hf$_2$O$_7$ and La$_2$Hf$_2$O$_7$, contributions from crystal field and phonon excitations could be distinguished (Fig.~\ref{Fig.1}).
Five crystal field excitations can be identified, as expected from the splitting of the $J=4$ intermediate coupling ground multiplet of a $4f^2$ ion in $D_{3d}$ local symmetry \footnote{In the absence of the crystal-field interaction the intermediate coupling basis states of Pr$^{3+}$ are dominated by the Hund's rule ground state $^3H_{4}$, with a small admixture of $^3F_{4}$ and $^1G_{4}$. See Ref.~\onlinecite{Boothroyd1992,Princep2013}.}
 into three doublets and three singlets with symmetry decomposition $3\Gamma_3^+ + 2\Gamma_1^+ + \Gamma_2^+$, see Ref.~\onlinecite{Boothroyd1992,Koster1963}.\footnote{In Koster's notation\cite{Koster1963}, the irreducible representations (irreps) $\Gamma_i$ are usually labelled such that the low-symmetry irreps have small $i$ indices. The `+' symbol appearing as a subscript to the index in $\Gamma_i^+$ indicates that the irrep is symmetric with respect to the inversion.}  The integrated intensities relative to the largest peak at 9.2 meV, obtained from fits to Gaussian peaks, are given in Table~\ref{TableI}.  The energies and intensities of these excitations were used to adjust the six parameters of the crystal field Hamiltonian 
\begin{IEEEeqnarray}{rCl} 
\hat{H}_{CF} &=& B_0^2C_0^2 + B_0^4C_0^4 + B_4^3(C_4^{-3} - C_4^{3}) + B_0^6C_0^6 \nonumber\\ 
&& +\: B_6^3(C_6^{-3} - C_6^{3}) + B_6^6(C_6^{-6} + C_6^{6}) \label{eq:cf_ham}
\end{IEEEeqnarray}

\noindent where $B_{q}^k$ denotes the crystal field parameters and $C_{q}^k$ are the components of the tensor operator $C^k$ (see Ref.~\onlinecite{Wybourne1965}), using the SPECTRE program and starting from the results obtained in a similar way for Pr$_2$Sn$_2$O$_7$~\cite{SPECTRE,Princep2013}.
We have used the complete set of 91 intermediate coupling basis states, which is required by the comparable strengths of the crystal field and spin orbit interactions in Pr. The refinement reaches the standard normalized goodness-of-fit parameter $\chi^2 = 1.2$. 
Table~\ref{TableI} gives the best-fit Wybourne coefficients $B_{q}^k$ together with the energies and relative intensities of the calculated levels. The temperature-dependence of the inverse magnetic susceptibility calculated from the set of fitted crystal-field parameters reproduces our experimental data (see inset of Fig.~\ref{Fig.1}\textbf{c}). The energy-level scheme is very similar to that determined for Pr$_2$Sn$_2$O$_7$\cite{Princep2013} or Pr$_2$Zr$_2$O$_7$\cite{Kimura:2013gj}, and the energy-level schemes of all three Pr$^{3+}$ pyrochlores have the same sequence of alternating doublets and singlets. In Pr$_2$Hf$_2$O$_7$ the ground state doublet wavefunction (symmetry $\Gamma_3^+$) written in terms of the $|^{2S+1}L_J,M_J\rangle$ basis is
\begin{IEEEeqnarray}{rCl} 
|\pm\rangle &=& 0.83 |^{3}H_4,\pm4\rangle \pm0.51 |^{3}H_4,\pm1\rangle -0.12 |^{3}H_4,\mp2\rangle  \nonumber\\ 
&& +\: 0.14 |^{1}G_4,\pm4\rangle \pm0.09 |^{1}G_4,\pm1\rangle  \nonumber\\
&& \mp \: 0.06 |^{3}H_5,\pm4\rangle \pm0.05 |^{3}H_5,\mp2\rangle.
\label{eq:wf}
\end{IEEEeqnarray}
\noindent We note the significant admixture of terms with 
$|M_{J} \neq \pm J\rangle$ which can increase quantum fluctuations~\cite{Onoda2010,Onoda2011}. Indeed, their existence implies additional effective transverse exchange terms in the spin ice Hamiltonian, which allows the ‘2-in-2-out’ configurations to tunnel among themselves. 
The calculated anisotropy in the susceptibility is $\chi_{\parallel} / \chi_{\perp}\sim 24$ at $T = 10$ K, where $\chi_{\parallel}$ and $\chi_{\perp}$ are the susceptibilities parallel and perpendicular to the $\langle111\rangle$ quantization axis of the crystal field. This ratio is about half of that in Pr$_2$Sn$_2$O$_7$ ~\cite{Princep2013}, in agreement with the reduction of the energy gap to the first excited level in Pr$_2$Hf$_2$O$_7$ ($\sim9$ meV \textit{vs.} $\sim18$ meV in Pr$_2$Sn$_2$O$_7$).

\begin{figure}
\includegraphics[width=8cm]{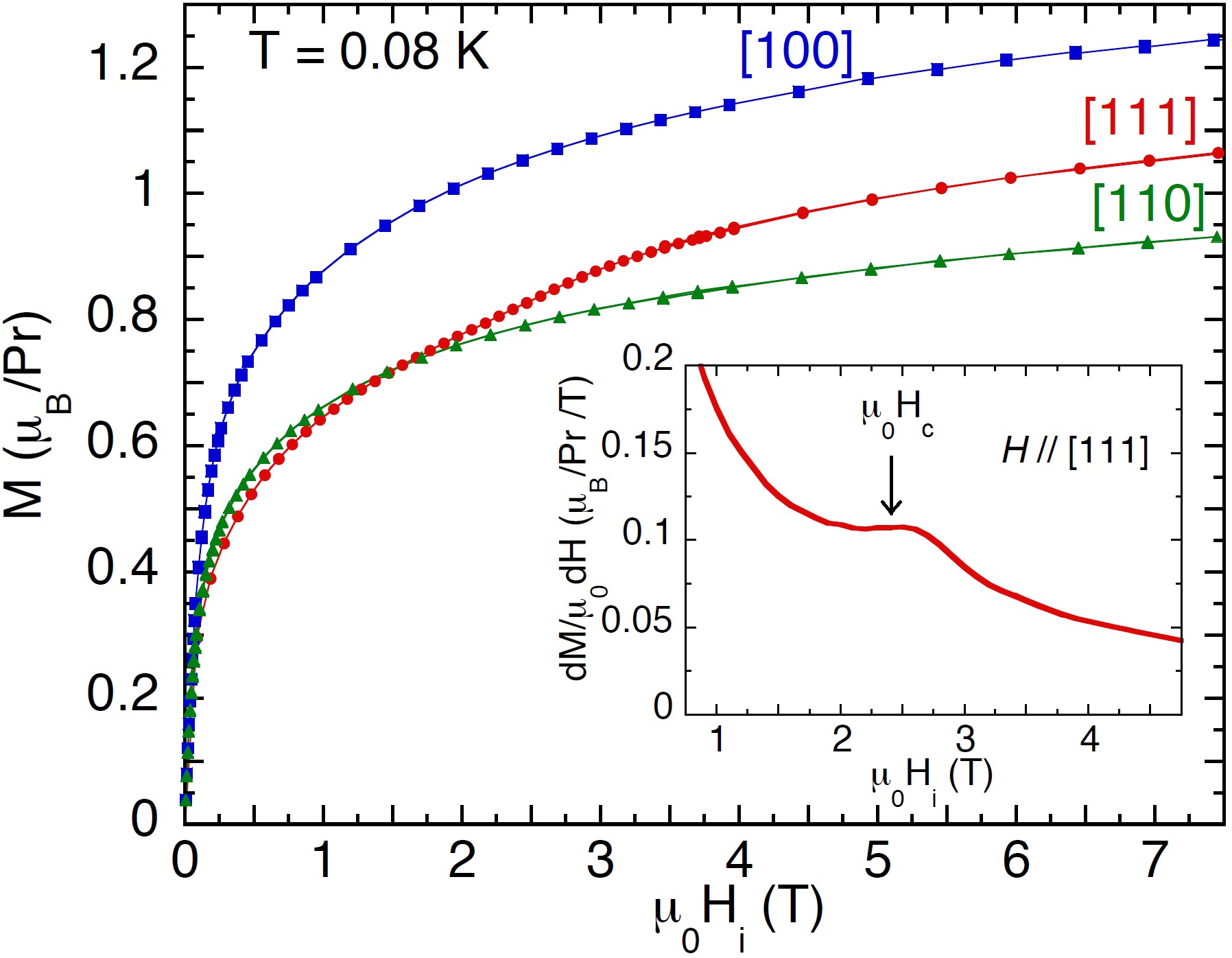}
\centering
\caption{Magnetization ($M$) as a function of internal magnetic field ($H_i$) at $T = 0.08$ K for $H$ applied along different crystallographic directions. Inset: derivative for $H\parallel [111]$, emphasizing the anomaly at $\mu_0H_c\sim2.4$ T.}
\label{Fig.2}
\end{figure}

The local anisotropy is confirmed by $M$ \textit{vs.} $H$ measurements along different crystallographic directions at $T=0.08$ K (Fig.~\ref{Fig.2}). The magnetization measured along [111], $M_{\rm [111]}$, is about $\sqrt{3}/2$ times $M_{\rm [100]}$, as expected for Ising spins on the pyrochlore lattice\cite{Harris1998}. $M_{[110]}/M_{[111]}$ is slightly larger than the predicted value of $2/\sqrt{6}$. The corresponding local Pr$^{3+}$ magnetic moment at $\mu_0H=8$~T is $2.2\pm0.1$~$\mu_{\rm B}/$Pr$^{3+}$. This is consistent with the magnetic moment $\mu_{\rm{Pr^{3+}}}=2.43$~$\mu_{\rm B}$ calculated from our ground doublet wavefunction (equation~\ref{eq:wf}) since the magnetization curves are not fully saturated at 8 T.

\subsection{Spin ice-like correlations}

\begin{figure}[b]
\includegraphics[width=\linewidth]{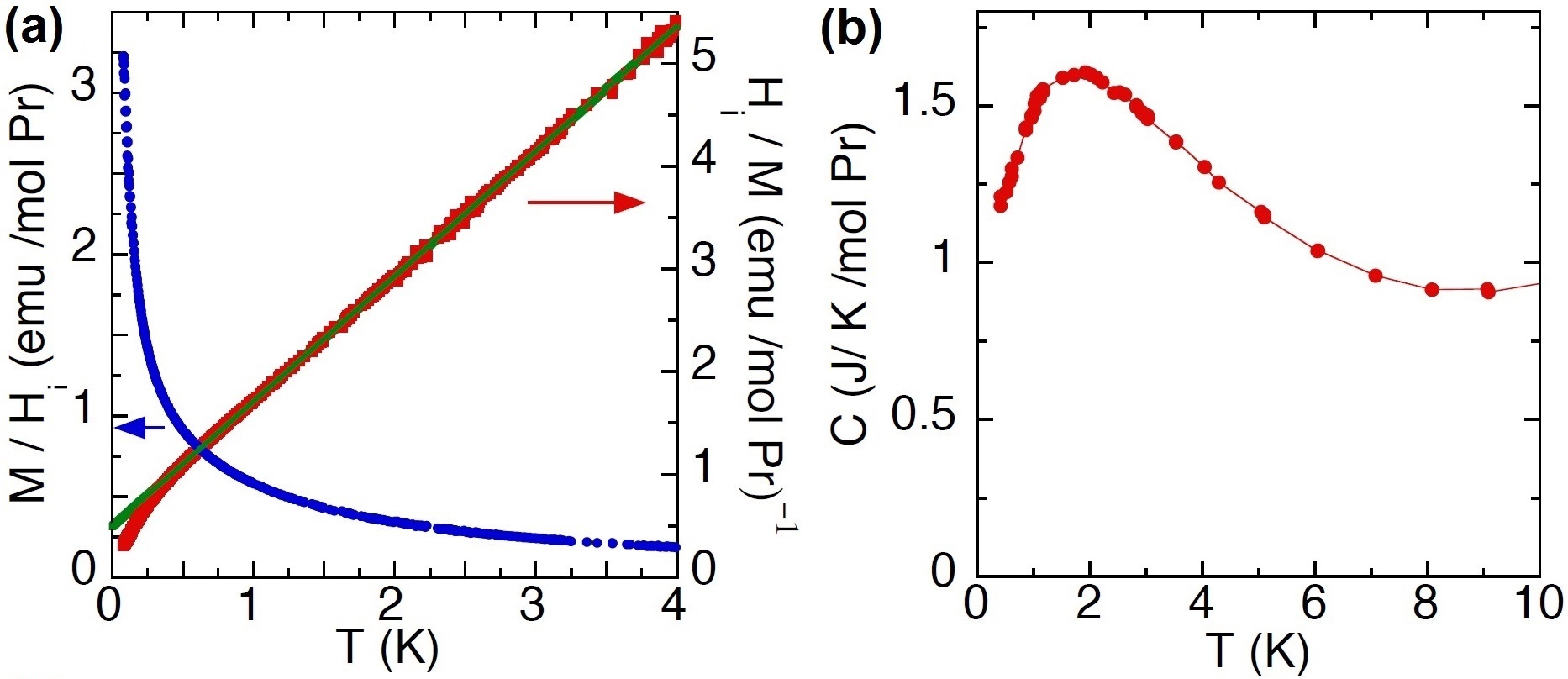}
\centering
\caption{(\textbf{a}) Magnetization ($M$) \textit{vs.} temperature ($T$) under a magnetic field $\mu_0H=4\times10^{-3}$ T applied along the [100] direction, presented as $M/H$ (blue data points) and $H/M$ (red data points). The green line is a fit to the Curie-Weiss law between $T=1$ and 4 K: $H/M=0.49 + 1.22 T$. (\textbf{b}) Heat capacity $C_p$ \textit{vs.} $T$.}
\label{Fig.3}
\end{figure}

At temperatures sufficiently low for the ground doublet to be isolated, the magnetic susceptibility $\chi~\sim~M/H$ (Fig.~\ref{Fig.3}\textbf{a}) follows a Curie-Weiss law indicating dominant antiferromagnetic interactions of the order of $|\theta_{CW}|=0.40\pm0.01$~K. A reasonable effective magnetic moment, $\mu_\mathrm{eff}=2.56\pm0.02$~$\mu_\mathrm{B}$,
is deduced from this linear fit ($1\leq~T\leq~4$~K). 
The temperature-dependence of $\chi$ demonstrates the absence of a magnetic transition down to 0.07~K. However, below $T=0.6$ K, $\chi$ exhibits a different behavior where it increases faster than the Curie-Weiss law as the temperature is lowered (c.f. $\chi^{-1}\sim H/M(T)$, Fig.~\ref{Fig.3}\textbf{a}). 
This is the signature of ferromagnetic-like correlations, which suggests, in the presence of such multiaxis Ising anisotropy, that the system is entering a spin ice-like state. 

The existence of such a local spin configuration is consistent with the anomaly observed in the magnetization curve at 0.08~K when the field is applied along the $[111]$ direction at $M \sim 0.82~\mu_{\rm B}/$Pr$^{3+}$ (Fig.~\ref{Fig.2}), very close to the value of $\mu_{\rm{Pr^{3+}}}/3=0.81~\mu_{\rm B}$. This anomaly, reminiscent of the kagome ice state observed in classical spin ice systems~\cite{Matsuhira2002}, is predicted in QSIs\cite{Molavian2009,Onoda2010,Onoda2011}, and has been reported in the metallic analogue Pr$_2$Ir$_2$O$_7$\cite{Machida2010}. It is a metamagnetic behavior associated with the transition from the `2-in-2-out' spin ice configuration to the `1-in-3-out' (`3-in-1-out') kagome ice state. The maximum of the derivative ${\rm d}M/{\rm d}H$ (Fig.~\ref{Fig.2}, inset) gives a critical field $\mu_0 H_c=2.4$~T which allows an estimate of the effective exchange energy scale in the system, $E_{\rm eff} = \frac{1}{3} \mu_0 \mu_{\rm eff} H_c/k_B \approx 1.2$~K. This is close to the value reported in the metallic Pr$^{3+}$ irridate, where the estimates of $\mu_0 H_c=2.3$~T and $\mu_{\rm{Pr^{3+}}}=2.7$~$\mu_{\rm B}$ lead to a slightly higher value of $E_{\rm eff} \approx 1.4$~K\cite{Machida2010}.

The heat capacity $C_p$ is characterized by a broad hump around $T=2$~K (Fig. \ref{Fig.3}\textbf{b}), reflecting the onset of magnetic correlations. 
Our data closely resembles those for Pr$_2$Zr$_2$O$_7$\cite{Matsuhira2009,Kimura:2013gj}, but the peak is slightly shifted towards lower temperature in the case of Pr$_2$Hf$_2$O$_7$.
We found that $C_p$ data extending far below the lower limit of our setup (0.3~K), for both Pr$_2$Hf$_2$O$_7$ and La$_2$Hf$_2$O$_7$, would be necessary for a proper integration procedure of $C_p(T)$ after subtraction of the lattice and hyperfine contributions. Therefore, here we do not provide the low temperature magnetic entropy vs. temperature of Pr$_2$Hf$_2$O$_7$.

\subsection{Spin dynamics from \textit{ac}-susceptibility}

\begin{figure}
\includegraphics[width=5.5cm]{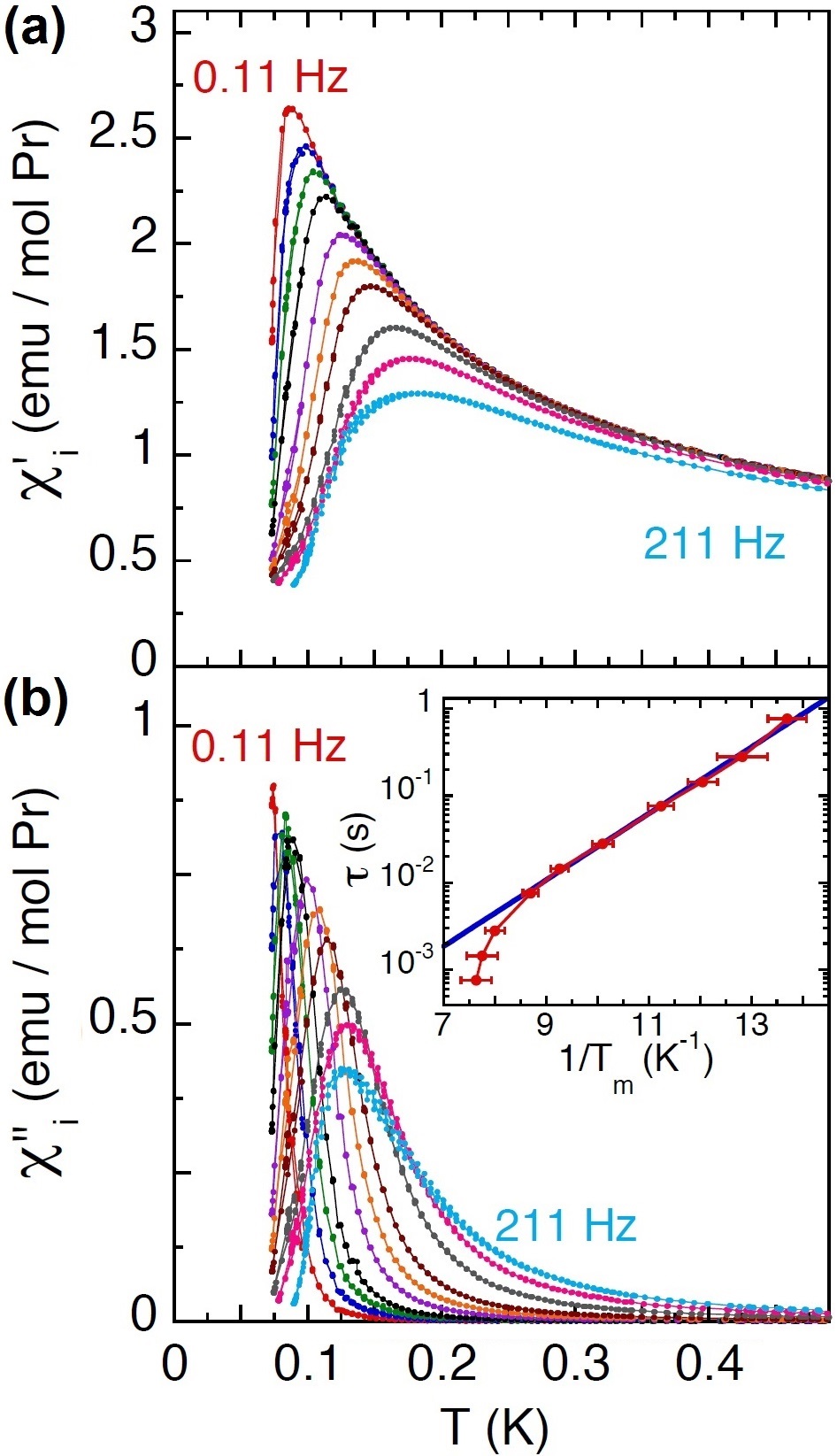}
\centering
\caption{Real part $\chi'_i$ (\textbf{a}) and imaginary part $\chi''_i$ (\textbf{b}) of the \textit{ac}-susceptibility \textit{vs.} $T$ at several frequencies $f$ between 0.11 and 211 Hz ($\mu_0H_{\rm ac}=5.47\times10^{-4}$~T, applied in an arbitrary direction). Inset: $\tau=1/2\pi f$ \textit{vs.} $1/T_{\rm peak}$, where $T_{\rm peak}$ is the temperature of the $\chi''_i$ maximum. The line is a fit to the Arrhenius law, for $\tau > 10^{-2}$s: $\tau= 4 \times 10^{-6} \exp(0.88/T_{\rm peak})$.}
\label{Fig.4}
\end{figure}

A frequency-dependent peak is observed below $T=0.2$~K in \textit{ac}-susceptibility experiments (Figure \ref{Fig.4}), i.e. deep in the correlated regime appearing at $T \lesssim 0.6$~K. This slowing down of the spin dynamics is rather similar to Pr$_2$Zr$_2$O$_7$\cite{Matsuhira2009,Kimura:2013gj}, but the amplitude of the out-of-phase part of the susceptibility $\chi''$ is larger, about one third of the in-phase susceptibility $\chi'$, meaning that a larger part of the magnetic moments is involved in the process. In comparison to the closely related zirconate and hafnate, the \textit{ac}-susceptibility signal in Pr$_2$Sn$_2$O$_7$ was reported to appear at slightly higher temperature, below $T=0.35$~K\cite{Matsuhira2002b}. For frequencies $f<20$ Hz the relaxation time probed in Pr$_2$Hf$_2$O$_7$ follows an Arrhenius law $\tau=\tau_0 \exp(E/k_BT)$ with an energy barrier $E \approx 0.86$ K, and a relatively long characteristic time $\tau_0 \approx 4 \times 10^{-6}$ s likely due to cooperative rather than single-ion dynamics. At higher temperature, the relaxation time deviates from this law and the peaks broaden, suggesting another regime with a distribution of relaxation times. 

The presence of a frequency-dependent peak in the \textit{ac}-susceptibility at low temperature is not only characteristic of the Pr$^{3+}$ zirconate~\cite{Matsuhira2009,Kimura:2013gj}, stannate~\cite{Matsuhira2002b} and hafnate (Figure \ref{Fig.4}) but it was also reported in the classical spin-ice materials. In these compounds, two regimes could be identified \cite{Snyder2004,Matsuhira2011,Quilliam2011,JaubertJPCM2011}. Upon cooling above $T \approx 5$~K, single-ion dynamics related to the anisotropy barrier are observed, firstly above the barrier, by thermal activation, and then across the barrier, via quantum tunneling. At lower temperature, $T \lesssim 5$~K, the dynamics slow down but persist due to monopole excitations above the spin ice ground state. In the Pr$^{3+}$ systems, fast single-ion dynamics might prevent the observation of the high temperature relaxation in the \textit{ac}-susceptibility, but we access the slow dynamics at very low temperature in the correlated state.

Finally, the \textit{ac}-susceptibility in Tb$_2$Ti$_2$O$_7$, a material with low-temperature correlations reminiscent of spin ices \cite{Fennell2012}, is also rather different compared to that in the Pr$^{3+}$ pyrochlores. Although the freezing appears below $T \approx 0.3$~K in this compound, the shape of the signals is rather unusual and their frequency-dependence does not follow an Arrhenius law. Instead, the peak shift may be described by a glassy behavior, using $\tau=\tau_0$~exp$[(E/T)^\sigma]$ with $E \approx 0.91$ K and $\tau_0 \approx 1.1 \times 10^{-9}$ s\cite{Lhotel2012}.

\subsection{Spin dynamics from neutron spectroscopy}

\begin{figure*}[ht]
\includegraphics[width=\linewidth]{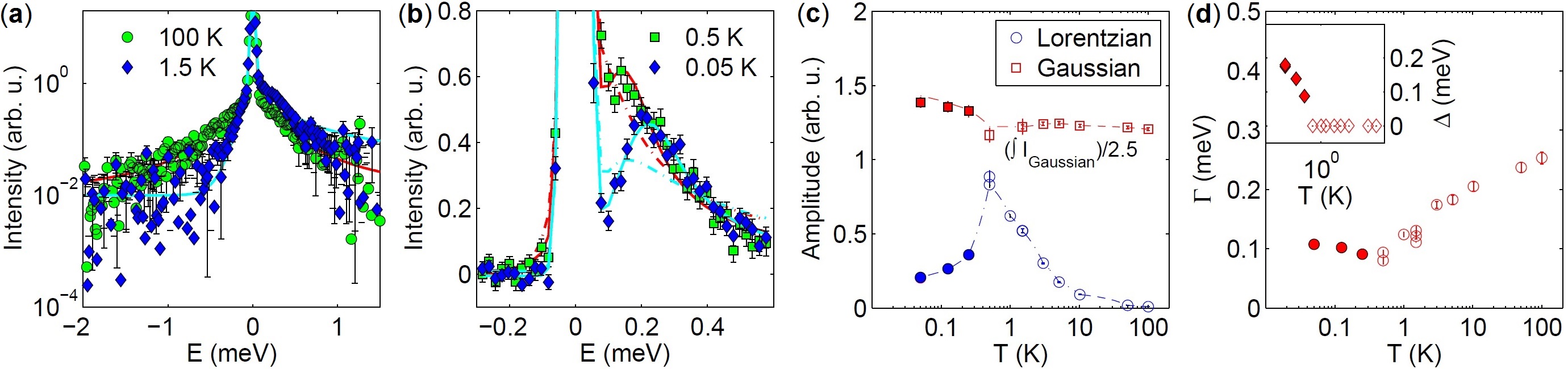}
\centering
\caption{Low-energy transfer neutron-scattering in the temperature range $0.05 \leq T \leq 100$~K. In \textbf{a} and \textbf{b} we present energy cuts integrated over $|Q|=[0.7,1.2]$~\AA$^{-1}$ after subtraction of the empty sample holder spectra. The lines are fit using the lineshape functions outlined in the text. In \textbf{b} we show two fits for each temperature: the inelastic scattering taken into account by either a quasielastic (dashed lines) or an inelastic (solid lines) Lorentzian contribution. In \textbf{c} we show the amplitude of the Lorentzian contribution and the integrated intensity of the Gaussian (divided by 2.5) as a function of $T$. The datasets are scaled to match the elastic intensity at 1.5 K. Open symbols show parameters derived from quasielastic fits, closed symbols from inelastic fits. Panel \textbf{d} gives the dependence of the width $\Gamma$ and energy center $\Delta$ (inset) of the Lorentzian contribution over the entire $T$ range.}
\label{Fig.5}
\end{figure*}

The spin dynamics were further investigated by low-energy neutron-scattering experiments. In Fig.~\ref{Fig.5}\textbf{a} and~\ref{Fig.5}\textbf{b} we present energy cuts integrated over $Q=[0.7,1.2]$~\AA$^{-1}$, after subtraction of the background of the empty sample holders. We fit the energy spectra using a Lorentzian peak shape weighted by detailed balance and convoluted with the instrumental resolution, plus a Gaussian peak at the elastic position\cite{Takatsu2012}.  The resolution is estimated from the width of the Gaussian (0.057 meV).  In the range  $0.5 \leq T \leq 100$ K, the data are well described by a quasielastic Lorentzian in combination with an elastic Gaussian (as in Fig.~\ref{Fig.5}\textbf{a}), but for $T<0.5$ K we have to allow the center of the Lorentzian to shift to a finite energy transfer $\Delta$. In Fig.~\ref{Fig.5}\textbf{b} we show that in this temperature range the quasielastic Lorentzian compares unfavorably with those centered at $E=\Delta$.

We interpret the quasielastic scattering observed at $0.5 \leq T \leq 100$ K as lifetime broadening of the single-ion ground state doublet. Upon cooling within this regime, the intrinsic width of the Lorentzian $\Gamma$ corrected for resolution decreases, while its amplitude increases (Fig.~\ref{Fig.5}\textbf{c} and~\ref{Fig.5}\textbf{d}), showing that the lifetime is increasing as the associated fluctuations are slowing. The characteristic times of the fluctuations probed by quasielastic neutron scattering within this regime are in the range from $\approx 5 \times 10^{-12}$ to $\approx 1 \times 10^{-11}$~s. Lifetime broadening of crystal field states is an uncorrelated, single ion fluctuation. In insulating compounds of rare-earth ions it is primarily due to interaction with phonons and population of excited states, but we were not able to fit the temperature dependence by a simple model such as an Arrhenius law or single Orbach process. It is interesting that these fluctuations are measurable down to 0.5~K in Pr$_2$Hf$_2$O$_7$. For example, in Ho$_2$Ti$_2$O$_7$, which also has a non-Kramers doublet ground state, single-ion fluctuations vanish at $T \approx 100$~K~\cite{Ehlers2002,Ruminy}, 200 times higher, despite the gap to the first excited state (22 meV) being only $\approx 2.4$ times greater than in Pr$_2$Hf$_2$O$_7$. Therefore, the overall observations suggest that another mechanism is at play in the case of Pr$_2$Hf$_2$O$_7$.

In the range $0.05 \leq T < 0.5$~K, the fluctuations in Pr$_2$Hf$_2$O$_7$ change from single-ion-like to some other type, and quasielastic spectral weight is redistributed. On one hand the elastic Gaussian increases in amplitude (Fig.~\ref{Fig.5}\textbf{c}), indicating the development of a small static component of the magnetic moment. The wave-vector dependence of this elastic signal cannot be appreciated from our experiment. On the other hand, a discrete excitation appears, whose energy center $\Delta$ increases to 0.2 meV $\sim2.3$~K (Fig.~\ref{Fig.5}\textbf{d}, inset), which we interpret~\footnote{Heat capacity measurements providing a proof for a vanishing spin entropy at the temperatures where the persistent fluctuations are observed using neutron spectroscopy would be necessary to claim the quantum origin of the fluctuations.} as a signature of cooperative quantum fluctuations\cite{Takatsu2012}.  In this respect, we might compare Pr$_2$Hf$_2$O$_7$ to Tb$_2$Ti$_2$O$_7$, where a correlated ground state is well known to exist\cite{Gardner:2010fu} and a shift of spectral weight similar to that described here also occurs\cite{Takatsu2012}. The appearance of an elastic component and a pronounced low-energy component in Pr$_2$Hf$_2$O$_7$ at $T < 0.5$~K, where $\chi(T)$ also deviates from its single-ion behavior, therefore qualitatively resembles the formation of a correlated state such as that in Tb$_2$Ti$_2$O$_7$.

\section{Conclusion}

The single-ion ground state of Pr$^{3+}$ in Pr$_2$Hf$_2$O$_7$ fulfills the requirements for the QSI state: small Ising-like moments and the potential for transverse fluctuations. Furthermore, we provide convincing arguments for the realization of a strongly-correlated state by observing \textit{(i)}~a discrete inelastic signal which appears below $T=0.5$~K, and \textit{(ii)}~slow fluctuations, probed by \textit{ac}-susceptibility, that appear in the same regime and may be associated with the manipulation of correlated objects appearing static for neutrons. 

Measurements under applied magnetic fields suggest that this strongly-correlated state is a QSI. Indeed, the critical field of the metamagnetic behavior is consistent with a transition from ‘2-in-2-out’ to ‘3-in-1-out’ configurations. The value of this metamagnetic field and the theoretical apparatus existing for Pr$^{3+}$ pyrochlores provide an estimate of the effective exchange energy in an insulating compound where the correlations that develop at low temperature are very probably related to a spin ice, and originate from exchange interactions. 

We suggest Pr$_2$Hf$_2$O$_7$ to be a model pyrochlore with thermally-isolated non-Kramers Ising moments and spin ice-like correlations.  Its availability as large, high quality, single crystals make it a promising system for investigating phenomena such as cooperative multipoles and chiral spin liquids\cite{Onoda2010}. Neutron scattering experiments on such samples are needed to examine the wave-vector dependence of the elastic and inelastic signals, and look for signs of the elusive photon modes and broadened pinch points predicted for QSIs.

\section{Acknowledgements}

We thank C. Paulsen for the use of his magnetometers and P. Lachkar for help with the heat capacity measurements; S. Petit for fruitful discussions; M. Bartkowiak and M. Zolliker for help with the dilution fridge experiments at SINQ; S. Rols for his careful assistance in setting up the IN4 spectrometer; and A. T. Boothroyd for the availability of  \textsc{SPECTRE}. We acknowledge funding from the European Community's Seventh Framework Program (grants 290605, COFUND: PSI-FELLOW; and 228464, Research Infrastructures under the FP7 Capacities Specific Programme, MICROKELVIN), and the Swiss National Science Foundation (Grants No. 200021\_140862, 200021\_138018, and 200020\_162626). The work at the University of Warwick was supported by EPSRC, UK (Grant EP/M028771/1). Neutron scattering experiments were carried out at the continuous spallation neutron source SINQ at the Paul Scherrer Institut at Villigen PSI in Switzerland and at the Institut Laue Langevin in Grenoble, France.


\begin{thebibliography}{59}%
\makeatletter
\providecommand \@ifxundefined [1]{%
 \@ifx{#1\undefined}
}%
\providecommand \@ifnum [1]{%
 \ifnum #1\expandafter \@firstoftwo
 \else \expandafter \@secondoftwo
 \fi
}%
\providecommand \@ifx [1]{%
 \ifx #1\expandafter \@firstoftwo
 \else \expandafter \@secondoftwo
 \fi
}%
\providecommand \natexlab [1]{#1}%
\providecommand \enquote  [1]{``#1''}%
\providecommand \bibnamefont  [1]{#1}%
\providecommand \bibfnamefont [1]{#1}%
\providecommand \citenamefont [1]{#1}%
\providecommand \href@noop [0]{\@secondoftwo}%
\providecommand \href [0]{\begingroup \@sanitize@url \@href}%
\providecommand \@href[1]{\@@startlink{#1}\@@href}%
\providecommand \@@href[1]{\endgroup#1\@@endlink}%
\providecommand \@sanitize@url [0]{\catcode `\\12\catcode `\$12\catcode
  `\&12\catcode `\#12\catcode `\^12\catcode `\_12\catcode `\%12\relax}%
\providecommand \@@startlink[1]{}%
\providecommand \@@endlink[0]{}%
\providecommand \url  [0]{\begingroup\@sanitize@url \@url }%
\providecommand \@url [1]{\endgroup\@href {#1}{\urlprefix }}%
\providecommand \urlprefix  [0]{URL }%
\providecommand \Eprint [0]{\href }%
\providecommand \doibase [0]{http://dx.doi.org/}%
\providecommand \selectlanguage [0]{\@gobble}%
\providecommand \bibinfo  [0]{\@secondoftwo}%
\providecommand \bibfield  [0]{\@secondoftwo}%
\providecommand \translation [1]{[#1]}%
\providecommand \BibitemOpen [0]{}%
\providecommand \bibitemStop [0]{}%
\providecommand \bibitemNoStop [0]{.\EOS\space}%
\providecommand \EOS [0]{\spacefactor3000\relax}%
\providecommand \BibitemShut  [1]{\csname bibitem#1\endcsname}%
\let\auto@bib@innerbib\@empty
\bibitem [{\citenamefont {Balents}(2010)}]{Balents:2010jx}%
  \BibitemOpen
  \bibfield  {author} {\bibinfo {author} {\bibfnamefont {L.}~\bibnamefont
  {Balents}},\ }\href@noop {} {\bibfield  {journal} {\bibinfo  {journal}
  {Nature}\ }\textbf {\bibinfo {volume} {464}},\ \bibinfo {pages} {199}
  (\bibinfo {year} {2010})}\BibitemShut {NoStop}%
\bibitem [{\citenamefont {Normand}(2009)}]{Normand:2009wk}%
  \BibitemOpen
  \bibfield  {author} {\bibinfo {author} {\bibfnamefont {B.}~\bibnamefont
  {Normand}},\ }\href@noop {} {\bibfield  {journal} {\bibinfo  {journal}
  {Contemp. Phys.}\ }\textbf {\bibinfo {volume} {50}},\ \bibinfo {pages} {533}
  (\bibinfo {year} {2009})}\BibitemShut {NoStop}%
\bibitem [{\citenamefont {Isakov}\ \emph {et~al.}(2011)\citenamefont {Isakov},
  \citenamefont {Hastings},\ and\ \citenamefont {Melko}}]{Isakov:2011fz}%
  \BibitemOpen
  \bibfield  {author} {\bibinfo {author} {\bibfnamefont {S.~V.}\ \bibnamefont
  {Isakov}}, \bibinfo {author} {\bibfnamefont {M.~B.}\ \bibnamefont
  {Hastings}}, \ and\ \bibinfo {author} {\bibfnamefont {R.~G.}\ \bibnamefont
  {Melko}},\ }\href@noop {} {\bibfield  {journal} {\bibinfo  {journal} {Nature
  Phys.}\ }\textbf {\bibinfo {volume} {7}},\ \bibinfo {pages} {772} (\bibinfo
  {year} {2011})}\BibitemShut {NoStop}%
\bibitem [{\citenamefont {Wen}(2002)}]{Wen:2002cy}%
  \BibitemOpen
  \bibfield  {author} {\bibinfo {author} {\bibfnamefont {X.-G.}\ \bibnamefont
  {Wen}},\ }\href@noop {} {\bibfield  {journal} {\bibinfo  {journal} {Phys.
  Rev. B}\ }\textbf {\bibinfo {volume} {65}},\ \bibinfo {pages} {165113}
  (\bibinfo {year} {2002})}\BibitemShut {NoStop}%
\bibitem [{\citenamefont {Gingras}\ and\ \citenamefont
  {McClarty}(2014)}]{Gingras:2014ip}%
  \BibitemOpen
  \bibfield  {author} {\bibinfo {author} {\bibfnamefont {M.~J.~P.}\
  \bibnamefont {Gingras}}\ and\ \bibinfo {author} {\bibfnamefont {P.~A.}\
  \bibnamefont {McClarty}},\ }\href@noop {} {\bibfield  {journal} {\bibinfo
  {journal} {Rep. Prog. Phys.}\ }\textbf {\bibinfo {volume} {77}},\ \bibinfo
  {pages} {056501} (\bibinfo {year} {2014})}\BibitemShut {NoStop}%
\bibitem [{\citenamefont {Moessner}\ and\ \citenamefont
  {Sondhi}(2010)}]{Moessner:2010ev}%
  \BibitemOpen
  \bibfield  {author} {\bibinfo {author} {\bibfnamefont {R.}~\bibnamefont
  {Moessner}}\ and\ \bibinfo {author} {\bibfnamefont {S.~L.}\ \bibnamefont
  {Sondhi}},\ }\href@noop {} {\bibfield  {journal} {\bibinfo  {journal} {Phys.
  Rev. Lett.}\ }\textbf {\bibinfo {volume} {105}},\ \bibinfo {pages} {166401}
  (\bibinfo {year} {2010})}\BibitemShut {NoStop}%
\bibitem [{\citenamefont {Gardner}\ \emph {et~al.}(2010)\citenamefont
  {Gardner}, \citenamefont {Gingras},\ and\ \citenamefont
  {Greedan}}]{Gardner:2010fu}%
  \BibitemOpen
  \bibfield  {author} {\bibinfo {author} {\bibfnamefont {J.~S.}\ \bibnamefont
  {Gardner}}, \bibinfo {author} {\bibfnamefont {M.~J.~P.}\ \bibnamefont
  {Gingras}}, \ and\ \bibinfo {author} {\bibfnamefont {J.~E.}\ \bibnamefont
  {Greedan}},\ }\href@noop {} {\bibfield  {journal} {\bibinfo  {journal} {Rev.
  Mod. Phys.}\ }\textbf {\bibinfo {volume} {82}},\ \bibinfo {pages} {53}
  (\bibinfo {year} {2010})}\BibitemShut {NoStop}%
\bibitem [{\citenamefont {Bramwell}\ and\ \citenamefont
  {Gingras}(2001)}]{Bramwell2001}%
  \BibitemOpen
  \bibfield  {author} {\bibinfo {author} {\bibfnamefont {S.~T.}\ \bibnamefont
  {Bramwell}}\ and\ \bibinfo {author} {\bibfnamefont {M.~J.~P.}\ \bibnamefont
  {Gingras}},\ }\href {\doibase 10.1126/science.1064761} {\bibfield  {journal}
  {\bibinfo  {journal} {Science}\ }\textbf {\bibinfo {volume} {294}},\ \bibinfo
  {pages} {1495} (\bibinfo {year} {2001})}\BibitemShut {NoStop}%
\bibitem [{\citenamefont {Bramwell}\ \emph {et~al.}(2001)\citenamefont
  {Bramwell}, \citenamefont {Harris}, \citenamefont {den Hertog}, \citenamefont
  {Gingras}, \citenamefont {Gardner}, \citenamefont {McMorrow}, \citenamefont
  {Wildes}, \citenamefont {Cornelius}, \citenamefont {Champion}, \citenamefont
  {Melko},\ and\ \citenamefont {Fennell}}]{Bramwell2001b}%
  \BibitemOpen
  \bibfield  {author} {\bibinfo {author} {\bibfnamefont {S.~T.}\ \bibnamefont
  {Bramwell}}, \bibinfo {author} {\bibfnamefont {M.~J.}\ \bibnamefont
  {Harris}}, \bibinfo {author} {\bibfnamefont {B.~C.}\ \bibnamefont {den
  Hertog}}, \bibinfo {author} {\bibfnamefont {M.~J.~P.}\ \bibnamefont
  {Gingras}}, \bibinfo {author} {\bibfnamefont {J.~S.}\ \bibnamefont
  {Gardner}}, \bibinfo {author} {\bibfnamefont {D.~F.}\ \bibnamefont
  {McMorrow}}, \bibinfo {author} {\bibfnamefont {A.~R.}\ \bibnamefont
  {Wildes}}, \bibinfo {author} {\bibfnamefont {A.~L.}\ \bibnamefont
  {Cornelius}}, \bibinfo {author} {\bibfnamefont {J.~D.~M.}\ \bibnamefont
  {Champion}}, \bibinfo {author} {\bibfnamefont {R.~G.}\ \bibnamefont {Melko}},
  \ and\ \bibinfo {author} {\bibfnamefont {T.}~\bibnamefont {Fennell}},\ }\href
  {\doibase 10.1103/PhysRevLett.87.047205} {\bibfield  {journal} {\bibinfo
  {journal} {Phys. Rev. Lett.}\ }\textbf {\bibinfo {volume} {87}},\ \bibinfo
  {pages} {047205} (\bibinfo {year} {2001})}\BibitemShut {NoStop}%
\bibitem [{\citenamefont {Castelnovo}\ \emph {et~al.}(2008)\citenamefont
  {Castelnovo}, \citenamefont {Moessner},\ and\ \citenamefont
  {Sondhi}}]{Castelnovo:2008hb}%
  \BibitemOpen
  \bibfield  {author} {\bibinfo {author} {\bibfnamefont {C.}~\bibnamefont
  {Castelnovo}}, \bibinfo {author} {\bibfnamefont {R.}~\bibnamefont
  {Moessner}}, \ and\ \bibinfo {author} {\bibfnamefont {S.~L.}\ \bibnamefont
  {Sondhi}},\ }\href@noop {} {\bibfield  {journal} {\bibinfo  {journal}
  {Nature}\ }\textbf {\bibinfo {volume} {451}},\ \bibinfo {pages} {42}
  (\bibinfo {year} {2008})}\BibitemShut {NoStop}%
\bibitem [{\citenamefont {Fennell}\ \emph {et~al.}(2009)\citenamefont
  {Fennell}, \citenamefont {Deen}, \citenamefont {Wildes}, \citenamefont
  {Schmalzl}, \citenamefont {Prabhakaran}, \citenamefont {Boothroyd},
  \citenamefont {Aldus}, \citenamefont {McMorrow},\ and\ \citenamefont
  {Bramwell}}]{Fennell2009}%
  \BibitemOpen
  \bibfield  {author} {\bibinfo {author} {\bibfnamefont {T.}~\bibnamefont
  {Fennell}}, \bibinfo {author} {\bibfnamefont {P.~P.}\ \bibnamefont {Deen}},
  \bibinfo {author} {\bibfnamefont {A.~R.}\ \bibnamefont {Wildes}}, \bibinfo
  {author} {\bibfnamefont {K.}~\bibnamefont {Schmalzl}}, \bibinfo {author}
  {\bibfnamefont {D.}~\bibnamefont {Prabhakaran}}, \bibinfo {author}
  {\bibfnamefont {A.~T.}\ \bibnamefont {Boothroyd}}, \bibinfo {author}
  {\bibfnamefont {R.~J.}\ \bibnamefont {Aldus}}, \bibinfo {author}
  {\bibfnamefont {D.~F.}\ \bibnamefont {McMorrow}}, \ and\ \bibinfo {author}
  {\bibfnamefont {S.~T.}\ \bibnamefont {Bramwell}},\ }\href {\doibase
  10.1126/science.1177582} {\bibfield  {journal} {\bibinfo  {journal}
  {Science}\ }\textbf {\bibinfo {volume} {326}},\ \bibinfo {pages} {415}
  (\bibinfo {year} {2009})}\BibitemShut {NoStop}%
\bibitem [{\citenamefont {Henley}(2010)}]{Henley:2010vo}%
  \BibitemOpen
  \bibfield  {author} {\bibinfo {author} {\bibfnamefont {C.~L.}\ \bibnamefont
  {Henley}},\ }\href@noop {} {\bibfield  {journal} {\bibinfo  {journal} {Annual
  Review of Condensed Matter Physics}\ }\textbf {\bibinfo {volume} {1}},\
  \bibinfo {pages} {179} (\bibinfo {year} {2010})}\BibitemShut {NoStop}%
\bibitem [{\citenamefont {Castelnovo}\ \emph {et~al.}(2012)\citenamefont
  {Castelnovo}, \citenamefont {Moessner},\ and\ \citenamefont
  {Sondhi}}]{Castelnovo:2012kk}%
  \BibitemOpen
  \bibfield  {author} {\bibinfo {author} {\bibfnamefont {C.}~\bibnamefont
  {Castelnovo}}, \bibinfo {author} {\bibfnamefont {R.}~\bibnamefont
  {Moessner}}, \ and\ \bibinfo {author} {\bibfnamefont {S.~L.}\ \bibnamefont
  {Sondhi}},\ }\href@noop {} {\bibfield  {journal} {\bibinfo  {journal} {Annual
  Review of Condensed Matter Physics}\ }\textbf {\bibinfo {volume} {3}},\
  \bibinfo {pages} {35} (\bibinfo {year} {2012})}\BibitemShut {NoStop}%
\bibitem [{\citenamefont {Curnoe}(2008)}]{Curnoe:2008gy}%
  \BibitemOpen
  \bibfield  {author} {\bibinfo {author} {\bibfnamefont {S.~H.}\ \bibnamefont
  {Curnoe}},\ }\href@noop {} {\bibfield  {journal} {\bibinfo  {journal} {Phys.
  Rev. B}\ }\textbf {\bibinfo {volume} {78}},\ \bibinfo {pages} {094418}
  (\bibinfo {year} {2008})}\BibitemShut {NoStop}%
\bibitem [{\citenamefont {Ross}\ \emph {et~al.}(2011)\citenamefont {Ross},
  \citenamefont {Savary}, \citenamefont {Gaulin},\ and\ \citenamefont
  {Balents}}]{Ross:2011tv}%
  \BibitemOpen
  \bibfield  {author} {\bibinfo {author} {\bibfnamefont {K.~A.}\ \bibnamefont
  {Ross}}, \bibinfo {author} {\bibfnamefont {L.}~\bibnamefont {Savary}},
  \bibinfo {author} {\bibfnamefont {B.~D.}\ \bibnamefont {Gaulin}}, \ and\
  \bibinfo {author} {\bibfnamefont {L.}~\bibnamefont {Balents}},\ }\href@noop
  {} {\bibfield  {journal} {\bibinfo  {journal} {Phys. Rev. X}\ }\textbf
  {\bibinfo {volume} {1}},\ \bibinfo {pages} {021002} (\bibinfo {year}
  {2011})}\BibitemShut {NoStop}%
\bibitem [{\citenamefont {Hermele}\ \emph {et~al.}(2004)\citenamefont
  {Hermele}, \citenamefont {Fisher},\ and\ \citenamefont
  {Balents}}]{Hermele:2004gg}%
  \BibitemOpen
  \bibfield  {author} {\bibinfo {author} {\bibfnamefont {M.}~\bibnamefont
  {Hermele}}, \bibinfo {author} {\bibfnamefont {M.~P.~A.}\ \bibnamefont
  {Fisher}}, \ and\ \bibinfo {author} {\bibfnamefont {L.}~\bibnamefont
  {Balents}},\ }\href@noop {} {\bibfield  {journal} {\bibinfo  {journal} {Phys.
  Rev. B}\ }\textbf {\bibinfo {volume} {69}},\ \bibinfo {pages} {064404}
  (\bibinfo {year} {2004})}\BibitemShut {NoStop}%
\bibitem [{\citenamefont {Savary}\ and\ \citenamefont
  {Balents}(2012)}]{Savary:2012cq}%
  \BibitemOpen
  \bibfield  {author} {\bibinfo {author} {\bibfnamefont {L.}~\bibnamefont
  {Savary}}\ and\ \bibinfo {author} {\bibfnamefont {L.}~\bibnamefont
  {Balents}},\ }\href@noop {} {\bibfield  {journal} {\bibinfo  {journal} {Phys.
  Rev. Lett.}\ }\textbf {\bibinfo {volume} {108}},\ \bibinfo {pages} {037202}
  (\bibinfo {year} {2012})}\BibitemShut {NoStop}%
\bibitem [{\citenamefont {Benton}\ \emph {et~al.}(2012)\citenamefont {Benton},
  \citenamefont {Sikora},\ and\ \citenamefont {Shannon}}]{Benton:2012ep}%
  \BibitemOpen
  \bibfield  {author} {\bibinfo {author} {\bibfnamefont {O.}~\bibnamefont
  {Benton}}, \bibinfo {author} {\bibfnamefont {O.}~\bibnamefont {Sikora}}, \
  and\ \bibinfo {author} {\bibfnamefont {N.}~\bibnamefont {Shannon}},\
  }\href@noop {} {\bibfield  {journal} {\bibinfo  {journal} {Phys. Rev. B}\
  }\textbf {\bibinfo {volume} {86}},\ \bibinfo {pages} {075154} (\bibinfo
  {year} {2012})}\BibitemShut {NoStop}%
\bibitem [{\citenamefont {Chang}\ \emph {et~al.}(2012)\citenamefont {Chang},
  \citenamefont {Onoda}, \citenamefont {Su}, \citenamefont {Kao}, \citenamefont
  {Tsuei}, \citenamefont {Yasui}, \citenamefont {Kakurai},\ and\ \citenamefont
  {Lees}}]{Chang:2012el}%
  \BibitemOpen
  \bibfield  {author} {\bibinfo {author} {\bibfnamefont {L.-J.}\ \bibnamefont
  {Chang}}, \bibinfo {author} {\bibfnamefont {S.}~\bibnamefont {Onoda}},
  \bibinfo {author} {\bibfnamefont {Y.}~\bibnamefont {Su}}, \bibinfo {author}
  {\bibfnamefont {Y.-J.}\ \bibnamefont {Kao}}, \bibinfo {author} {\bibfnamefont
  {K.-D.}\ \bibnamefont {Tsuei}}, \bibinfo {author} {\bibfnamefont
  {Y.}~\bibnamefont {Yasui}}, \bibinfo {author} {\bibfnamefont
  {K.}~\bibnamefont {Kakurai}}, \ and\ \bibinfo {author} {\bibfnamefont
  {M.~R.}\ \bibnamefont {Lees}},\ }\href@noop {} {\bibfield  {journal}
  {\bibinfo  {journal} {Nat. Commun.}\ }\textbf {\bibinfo {volume} {3}},\
  \bibinfo {pages} {992} (\bibinfo {year} {2012})}\BibitemShut {NoStop}%
\bibitem [{\citenamefont {Applegate}\ \emph {et~al.}(2012)\citenamefont
  {Applegate}, \citenamefont {Hayre}, \citenamefont {Singh}, \citenamefont
  {Lin}, \citenamefont {Day},\ and\ \citenamefont {Gingras}}]{Applegate2012}%
  \BibitemOpen
  \bibfield  {author} {\bibinfo {author} {\bibfnamefont {R.}~\bibnamefont
  {Applegate}}, \bibinfo {author} {\bibfnamefont {N.~R.}\ \bibnamefont
  {Hayre}}, \bibinfo {author} {\bibfnamefont {R.~R.~P.}\ \bibnamefont {Singh}},
  \bibinfo {author} {\bibfnamefont {T.}~\bibnamefont {Lin}}, \bibinfo {author}
  {\bibfnamefont {A.~G.~R.}\ \bibnamefont {Day}}, \ and\ \bibinfo {author}
  {\bibfnamefont {M.~J.~P.}\ \bibnamefont {Gingras}},\ }\href {\doibase
  10.1103/PhysRevLett.109.097205} {\bibfield  {journal} {\bibinfo  {journal}
  {Phys. Rev. Lett.}\ }\textbf {\bibinfo {volume} {109}},\ \bibinfo {pages}
  {097205} (\bibinfo {year} {2012})}\BibitemShut {NoStop}%
\bibitem [{\citenamefont {Robert}\ \emph {et~al.}(2015)\citenamefont {Robert},
  \citenamefont {Lhotel}, \citenamefont {Remenyi}, \citenamefont {Sahling},
  \citenamefont {Mirebeau}, \citenamefont {Decorse}, \citenamefont {Canals},\
  and\ \citenamefont {Petit}}]{Robert2015}%
  \BibitemOpen
  \bibfield  {author} {\bibinfo {author} {\bibfnamefont {J.}~\bibnamefont
  {Robert}}, \bibinfo {author} {\bibfnamefont {E.}~\bibnamefont {Lhotel}},
  \bibinfo {author} {\bibfnamefont {G.}~\bibnamefont {Remenyi}}, \bibinfo
  {author} {\bibfnamefont {S.}~\bibnamefont {Sahling}}, \bibinfo {author}
  {\bibfnamefont {I.}~\bibnamefont {Mirebeau}}, \bibinfo {author}
  {\bibfnamefont {C.}~\bibnamefont {Decorse}}, \bibinfo {author} {\bibfnamefont
  {B.}~\bibnamefont {Canals}}, \ and\ \bibinfo {author} {\bibfnamefont
  {S.}~\bibnamefont {Petit}},\ }\href {\doibase 10.1103/PhysRevB.92.064425}
  {\bibfield  {journal} {\bibinfo  {journal} {Phys. Rev. B}\ }\textbf {\bibinfo
  {volume} {92}},\ \bibinfo {pages} {064425} (\bibinfo {year}
  {2015})}\BibitemShut {NoStop}%
\bibitem [{\citenamefont {Jaubert}\ \emph {et~al.}(2015)\citenamefont
  {Jaubert}, \citenamefont {Benton}, \citenamefont {Rau}, \citenamefont
  {Oitmaa}, \citenamefont {Singh}, \citenamefont {Shannon},\ and\ \citenamefont
  {Gingras}}]{Jaubert2015}%
  \BibitemOpen
  \bibfield  {author} {\bibinfo {author} {\bibfnamefont {L.~D.~C.}\
  \bibnamefont {Jaubert}}, \bibinfo {author} {\bibfnamefont {O.}~\bibnamefont
  {Benton}}, \bibinfo {author} {\bibfnamefont {J.~G.}\ \bibnamefont {Rau}},
  \bibinfo {author} {\bibfnamefont {J.}~\bibnamefont {Oitmaa}}, \bibinfo
  {author} {\bibfnamefont {R.~R.~P.}\ \bibnamefont {Singh}}, \bibinfo {author}
  {\bibfnamefont {N.}~\bibnamefont {Shannon}}, \ and\ \bibinfo {author}
  {\bibfnamefont {M.~J.~P.}\ \bibnamefont {Gingras}},\ }\href {\doibase
  10.1103/PhysRevLett.115.267208} {\bibfield  {journal} {\bibinfo  {journal}
  {Phys. Rev. Lett.}\ }\textbf {\bibinfo {volume} {115}},\ \bibinfo {pages}
  {267208} (\bibinfo {year} {2015})}\BibitemShut {NoStop}%
\bibitem [{\citenamefont {Zhou}\ \emph {et~al.}(2008)\citenamefont {Zhou},
  \citenamefont {Wiebe}, \citenamefont {Janik}, \citenamefont {Balicas},
  \citenamefont {Yo}, \citenamefont {Qiu}, \citenamefont {Copley},\ and\
  \citenamefont {Gardner}}]{Zhou:2008cz}%
  \BibitemOpen
  \bibfield  {author} {\bibinfo {author} {\bibfnamefont {H.~D.}\ \bibnamefont
  {Zhou}}, \bibinfo {author} {\bibfnamefont {C.~R.}\ \bibnamefont {Wiebe}},
  \bibinfo {author} {\bibfnamefont {J.~A.}\ \bibnamefont {Janik}}, \bibinfo
  {author} {\bibfnamefont {L.}~\bibnamefont {Balicas}}, \bibinfo {author}
  {\bibfnamefont {Y.~J.}\ \bibnamefont {Yo}}, \bibinfo {author} {\bibfnamefont
  {Y.}~\bibnamefont {Qiu}}, \bibinfo {author} {\bibfnamefont {J.~R.~D.}\
  \bibnamefont {Copley}}, \ and\ \bibinfo {author} {\bibfnamefont {J.~S.}\
  \bibnamefont {Gardner}},\ }\href@noop {} {\bibfield  {journal} {\bibinfo
  {journal} {Phys. Rev. Lett.}\ }\textbf {\bibinfo {volume} {101}},\ \bibinfo
  {pages} {227204} (\bibinfo {year} {2008})}\BibitemShut {NoStop}%
\bibitem [{\citenamefont {Matsuhira}\ \emph {et~al.}(2009)\citenamefont
  {Matsuhira}, \citenamefont {Sekine}, \citenamefont {Paulsen}, \citenamefont
  {Wakeshima}, \citenamefont {Hinatsu}, \citenamefont {Kitazawa}, \citenamefont
  {Kiuchi}, \citenamefont {Hiroi},\ and\ \citenamefont
  {Takagi}}]{Matsuhira2009}%
  \BibitemOpen
  \bibfield  {author} {\bibinfo {author} {\bibfnamefont {K.}~\bibnamefont
  {Matsuhira}}, \bibinfo {author} {\bibfnamefont {C.}~\bibnamefont {Sekine}},
  \bibinfo {author} {\bibfnamefont {C.}~\bibnamefont {Paulsen}}, \bibinfo
  {author} {\bibfnamefont {M.}~\bibnamefont {Wakeshima}}, \bibinfo {author}
  {\bibfnamefont {Y.}~\bibnamefont {Hinatsu}}, \bibinfo {author} {\bibfnamefont
  {T.}~\bibnamefont {Kitazawa}}, \bibinfo {author} {\bibfnamefont
  {Y.}~\bibnamefont {Kiuchi}}, \bibinfo {author} {\bibfnamefont
  {Z.}~\bibnamefont {Hiroi}}, \ and\ \bibinfo {author} {\bibfnamefont
  {S.}~\bibnamefont {Takagi}},\ }\href@noop {} {\bibfield  {journal} {\bibinfo
  {journal} {J. Phys. Conf. Series}\ }\textbf {\bibinfo {volume} {145}},\
  \bibinfo {pages} {012031} (\bibinfo {year} {2009})}\BibitemShut {NoStop}%
\bibitem [{\citenamefont {Onoda}\ and\ \citenamefont
  {Tanaka}(2010)}]{Onoda2010}%
  \BibitemOpen
  \bibfield  {author} {\bibinfo {author} {\bibfnamefont {S.}~\bibnamefont
  {Onoda}}\ and\ \bibinfo {author} {\bibfnamefont {Y.}~\bibnamefont {Tanaka}},\
  }\href {\doibase 10.1103/PhysRevLett.105.047201} {\bibfield  {journal}
  {\bibinfo  {journal} {Phys. Rev. Lett.}\ }\textbf {\bibinfo {volume} {105}},\
  \bibinfo {pages} {047201} (\bibinfo {year} {2010})}\BibitemShut {NoStop}%
\bibitem [{\citenamefont {Lee}\ \emph {et~al.}(2012)\citenamefont {Lee},
  \citenamefont {Onoda},\ and\ \citenamefont {Balents}}]{Lee2012}%
  \BibitemOpen
  \bibfield  {author} {\bibinfo {author} {\bibfnamefont {S.~B.}\ \bibnamefont
  {Lee}}, \bibinfo {author} {\bibfnamefont {S.}~\bibnamefont {Onoda}}, \ and\
  \bibinfo {author} {\bibfnamefont {L.}~\bibnamefont {Balents}},\ }\href
  {\doibase 10.1103/PhysRevB.86.104412} {\bibfield  {journal} {\bibinfo
  {journal} {Phys. Rev. B}\ }\textbf {\bibinfo {volume} {86}},\ \bibinfo
  {pages} {104412} (\bibinfo {year} {2012})}\BibitemShut {NoStop}%
\bibitem [{\citenamefont {Kimura}\ \emph {et~al.}(2013)\citenamefont {Kimura},
  \citenamefont {Nakatsuji}, \citenamefont {Wen}, \citenamefont {Broholm},
  \citenamefont {Stone}, \citenamefont {Nishibori},\ and\ \citenamefont
  {Sawa}}]{Kimura:2013gj}%
  \BibitemOpen
  \bibfield  {author} {\bibinfo {author} {\bibfnamefont {K.}~\bibnamefont
  {Kimura}}, \bibinfo {author} {\bibfnamefont {S.}~\bibnamefont {Nakatsuji}},
  \bibinfo {author} {\bibfnamefont {J.-J.}\ \bibnamefont {Wen}}, \bibinfo
  {author} {\bibfnamefont {C.}~\bibnamefont {Broholm}}, \bibinfo {author}
  {\bibfnamefont {M.~B.}\ \bibnamefont {Stone}}, \bibinfo {author}
  {\bibfnamefont {E.}~\bibnamefont {Nishibori}}, \ and\ \bibinfo {author}
  {\bibfnamefont {H.}~\bibnamefont {Sawa}},\ }\href@noop {} {\bibfield
  {journal} {\bibinfo  {journal} {Nat. Commun.}\ }\textbf {\bibinfo {volume}
  {4}},\ \bibinfo {pages} {1934} (\bibinfo {year} {2013})}\BibitemShut
  {NoStop}%
\bibitem [{\citenamefont {Sibille}\ \emph {et~al.}(2015)\citenamefont
  {Sibille}, \citenamefont {Lhotel}, \citenamefont {Pomjakushin}, \citenamefont
  {Baines}, \citenamefont {Fennell},\ and\ \citenamefont
  {Kenzelmann}}]{Sibille2015}%
  \BibitemOpen
  \bibfield  {author} {\bibinfo {author} {\bibfnamefont {R.}~\bibnamefont
  {Sibille}}, \bibinfo {author} {\bibfnamefont {E.}~\bibnamefont {Lhotel}},
  \bibinfo {author} {\bibfnamefont {V.}~\bibnamefont {Pomjakushin}}, \bibinfo
  {author} {\bibfnamefont {C.}~\bibnamefont {Baines}}, \bibinfo {author}
  {\bibfnamefont {T.}~\bibnamefont {Fennell}}, \ and\ \bibinfo {author}
  {\bibfnamefont {M.}~\bibnamefont {Kenzelmann}},\ }\href {\doibase
  10.1103/PhysRevLett.115.097202} {\bibfield  {journal} {\bibinfo  {journal}
  {Phys. Rev. Lett.}\ }\textbf {\bibinfo {volume} {115}},\ \bibinfo {pages}
  {097202} (\bibinfo {year} {2015})}\BibitemShut {NoStop}%
\bibitem [{\citenamefont {Craig}(2009)}]{Craig2009}%
  \BibitemOpen
  \bibfield  {author} {\bibinfo {author} {\bibfnamefont {H.~A.}\ \bibnamefont
  {Craig}},\ }\href@noop {} {\bibfield  {journal} {\bibinfo  {journal}
  {Explorations: An Undergraduate Research Journal}\ ,\ \bibinfo {pages} {23}}
  (\bibinfo {year} {2009})}\BibitemShut {NoStop}%
\bibitem [{\citenamefont {Anand}\ \emph {et~al.}(2015)\citenamefont {Anand},
  \citenamefont {Bera}, \citenamefont {Xu}, \citenamefont {Herrmannsd\"orfer},
  \citenamefont {Ritter},\ and\ \citenamefont {Lake}}]{Anand2015}%
  \BibitemOpen
  \bibfield  {author} {\bibinfo {author} {\bibfnamefont {V.~K.}\ \bibnamefont
  {Anand}}, \bibinfo {author} {\bibfnamefont {A.~K.}\ \bibnamefont {Bera}},
  \bibinfo {author} {\bibfnamefont {J.}~\bibnamefont {Xu}}, \bibinfo {author}
  {\bibfnamefont {T.}~\bibnamefont {Herrmannsd\"orfer}}, \bibinfo {author}
  {\bibfnamefont {C.}~\bibnamefont {Ritter}}, \ and\ \bibinfo {author}
  {\bibfnamefont {B.}~\bibnamefont {Lake}},\ }\href {\doibase
  10.1103/PhysRevB.92.184418} {\bibfield  {journal} {\bibinfo  {journal} {Phys.
  Rev. B}\ }\textbf {\bibinfo {volume} {92}},\ \bibinfo {pages} {184418}
  (\bibinfo {year} {2015})}\BibitemShut {NoStop}%
\bibitem [{\citenamefont {Ciomaga~Hatnean}\ \emph {et~al.}(2015)\citenamefont
  {Ciomaga~Hatnean}, \citenamefont {Sibille}, \citenamefont {Lees},
  \citenamefont {Kenzelmann}, \citenamefont {Ban}, \citenamefont
  {Pomjakushin},\ and\ \citenamefont {Balakrishnan}}]{Ciomaga2015}%
  \BibitemOpen
  \bibfield  {author} {\bibinfo {author} {\bibfnamefont {M.}~\bibnamefont
  {Ciomaga~Hatnean}}, \bibinfo {author} {\bibfnamefont {R.}~\bibnamefont
  {Sibille}}, \bibinfo {author} {\bibfnamefont {M.~R.}\ \bibnamefont {Lees}},
  \bibinfo {author} {\bibfnamefont {M.}~\bibnamefont {Kenzelmann}}, \bibinfo
  {author} {\bibfnamefont {V.}~\bibnamefont {Ban}}, \bibinfo {author}
  {\bibfnamefont {V.}~\bibnamefont {Pomjakushin}}, \ and\ \bibinfo {author}
  {\bibfnamefont {G.}~\bibnamefont {Balakrishnan}},\ }\href@noop {} {\bibfield
  {journal} {\bibinfo  {journal} {in preparation}\ } (\bibinfo {year}
  {2015})}\BibitemShut {NoStop}%
\bibitem [{\citenamefont {Shevchenko}\ \emph {et~al.}(1984)\citenamefont
  {Shevchenko}, \citenamefont {Lopato},\ and\ \citenamefont
  {Nazarenko}}]{Shevchenko1984}%
  \BibitemOpen
  \bibfield  {author} {\bibinfo {author} {\bibfnamefont {A.~V.}\ \bibnamefont
  {Shevchenko}}, \bibinfo {author} {\bibfnamefont {L.~M.}\ \bibnamefont
  {Lopato}}, \ and\ \bibinfo {author} {\bibfnamefont {L.~V.}\ \bibnamefont
  {Nazarenko}},\ }\href@noop {} {\bibfield  {journal} {\bibinfo  {journal}
  {Inorg. Mater.}\ }\textbf {\bibinfo {volume} {20}},\ \bibinfo {pages} {1615}
  (\bibinfo {year} {1984})}\BibitemShut {NoStop}%
\bibitem [{\citenamefont {Karthik}\ \emph {et~al.}(2012)\citenamefont
  {Karthik}, \citenamefont {Anderson}, \citenamefont {Gout},\ and\
  \citenamefont {Ubic}}]{Karthik2012}%
  \BibitemOpen
  \bibfield  {author} {\bibinfo {author} {\bibfnamefont {C.}~\bibnamefont
  {Karthik}}, \bibinfo {author} {\bibfnamefont {T.~J.}\ \bibnamefont
  {Anderson}}, \bibinfo {author} {\bibfnamefont {D.}~\bibnamefont {Gout}}, \
  and\ \bibinfo {author} {\bibfnamefont {R.}~\bibnamefont {Ubic}},\ }\href@noop
  {} {\bibfield  {journal} {\bibinfo  {journal} {J. Solid State Chem.}\
  }\textbf {\bibinfo {volume} {194}},\ \bibinfo {pages} {168} (\bibinfo {year}
  {2012})}\BibitemShut {NoStop}%
\bibitem [{\citenamefont {Blanchard}\ \emph {et~al.}(2013)\citenamefont
  {Blanchard}, \citenamefont {Liu}, \citenamefont {Kennedy}, \citenamefont
  {Ling}, \citenamefont {Avdeev}, \citenamefont {Aitken}, \citenamefont
  {Cowie},\ and\ \citenamefont {Tadich}}]{Blanchard2013}%
  \BibitemOpen
  \bibfield  {author} {\bibinfo {author} {\bibfnamefont {P.~E.~R.}\
  \bibnamefont {Blanchard}}, \bibinfo {author} {\bibfnamefont {S.}~\bibnamefont
  {Liu}}, \bibinfo {author} {\bibfnamefont {B.~J.}\ \bibnamefont {Kennedy}},
  \bibinfo {author} {\bibfnamefont {C.~D.}\ \bibnamefont {Ling}}, \bibinfo
  {author} {\bibfnamefont {M.}~\bibnamefont {Avdeev}}, \bibinfo {author}
  {\bibfnamefont {J.~B.}\ \bibnamefont {Aitken}}, \bibinfo {author}
  {\bibfnamefont {B.~C.~C.}\ \bibnamefont {Cowie}}, \ and\ \bibinfo {author}
  {\bibfnamefont {A.}~\bibnamefont {Tadich}},\ }\href@noop {} {\bibfield
  {journal} {\bibinfo  {journal} {J. Phys. Chem. C}\ }\textbf {\bibinfo
  {volume} {117}},\ \bibinfo {pages} {2266} (\bibinfo {year}
  {2013})}\BibitemShut {NoStop}%
\bibitem [{\citenamefont {Paulsen}(2001)}]{Paulsen01}%
  \BibitemOpen
  \bibfield  {author} {\bibinfo {author} {\bibfnamefont {C.}~\bibnamefont
  {Paulsen}},\ }in\ \href@noop {} {\emph {\bibinfo {booktitle} {Introduction to
  Physical Techniques in Molecular Magnetism: Structural and Macroscopic
  Techniques - Yesa 1999}}},\ \bibinfo {editor} {edited by\ \bibinfo {editor}
  {\bibfnamefont {F.}~\bibnamefont {Palacio}}, \bibinfo {editor} {\bibfnamefont
  {E.}~\bibnamefont {Ressouche}}, \ and\ \bibinfo {editor} {\bibfnamefont
  {J.}~\bibnamefont {Schweizer}}}\ (\bibinfo  {publisher} {Servicio de
  Publicaciones de la Universidad de Zaragoza},\ \bibinfo {address}
  {Zaragoza},\ \bibinfo {year} {2001})\ p.~\bibinfo {pages} {1}\BibitemShut
  {NoStop}%
\bibitem [{\citenamefont {Aharoni}(1998)}]{Aharoni1998}%
  \BibitemOpen
  \bibfield  {author} {\bibinfo {author} {\bibfnamefont {A.}~\bibnamefont
  {Aharoni}},\ }\href {\doibase http://dx.doi.org/10.1063/1.367113} {\bibfield
  {journal} {\bibinfo  {journal} {J. Appl. Phys.}\ }\textbf {\bibinfo {volume}
  {83}},\ \bibinfo {pages} {3432} (\bibinfo {year} {1998})}\BibitemShut
  {NoStop}%
\bibitem [{\citenamefont {Princep}\ \emph {et~al.}(2013)\citenamefont
  {Princep}, \citenamefont {Prabhakaran}, \citenamefont {Boothroyd},\ and\
  \citenamefont {Adroja}}]{Princep2013}%
  \BibitemOpen
  \bibfield  {author} {\bibinfo {author} {\bibfnamefont {A.~J.}\ \bibnamefont
  {Princep}}, \bibinfo {author} {\bibfnamefont {D.}~\bibnamefont
  {Prabhakaran}}, \bibinfo {author} {\bibfnamefont {A.~T.}\ \bibnamefont
  {Boothroyd}}, \ and\ \bibinfo {author} {\bibfnamefont {D.~T.}\ \bibnamefont
  {Adroja}},\ }\href@noop {} {\bibfield  {journal} {\bibinfo  {journal} {Phys.
  Rev. B}\ }\textbf {\bibinfo {volume} {88}},\ \bibinfo {pages} {104421}
  (\bibinfo {year} {2013})}\BibitemShut {NoStop}%
\bibitem [{Note1()}]{Note1}%
  \BibitemOpen
  \bibinfo {note} {In the absence of the crystal-field interaction the
  intermediate coupling basis states of Pr$^{3+}$ are dominated by the Hund's
  rule ground state $^3H_{4}$, with a small admixture of $^3F_{4}$ and
  $^1G_{4}$. See Ref.~\protect \rev@citealpnum
  {Boothroyd1992,Princep2013}.}\BibitemShut {Stop}%
\bibitem [{\citenamefont {Boothroyd}\ \emph {et~al.}(1992)\citenamefont
  {Boothroyd}, \citenamefont {Doyle}, \citenamefont {Paul},\ and\ \citenamefont
  {Osborn}}]{Boothroyd1992}%
  \BibitemOpen
  \bibfield  {author} {\bibinfo {author} {\bibfnamefont {A.~T.}\ \bibnamefont
  {Boothroyd}}, \bibinfo {author} {\bibfnamefont {S.~M.}\ \bibnamefont
  {Doyle}}, \bibinfo {author} {\bibfnamefont {D.~M.}\ \bibnamefont {Paul}}, \
  and\ \bibinfo {author} {\bibfnamefont {R.}~\bibnamefont {Osborn}},\ }\href
  {\doibase 10.1103/PhysRevB.45.10075} {\bibfield  {journal} {\bibinfo
  {journal} {Phys. Rev. B}\ }\textbf {\bibinfo {volume} {45}},\ \bibinfo
  {pages} {10075} (\bibinfo {year} {1992})}\BibitemShut {NoStop}%
\bibitem [{\citenamefont {Koster}\ \emph {et~al.}(1963)\citenamefont {Koster},
  \citenamefont {Dimmock}, \citenamefont {Wheeler},\ and\ \citenamefont
  {Statz}}]{Koster1963}%
  \BibitemOpen
  \bibfield  {author} {\bibinfo {author} {\bibfnamefont {G.~F.}\ \bibnamefont
  {Koster}}, \bibinfo {author} {\bibfnamefont {J.}~\bibnamefont {Dimmock}},
  \bibinfo {author} {\bibfnamefont {R.}~\bibnamefont {Wheeler}}, \ and\
  \bibinfo {author} {\bibfnamefont {H.}~\bibnamefont {Statz}},\ }in\ \href@noop
  {} {\emph {\bibinfo {booktitle} {Properties of the thirty-two point
  groups}}}\ (\bibinfo  {publisher} {MIT Press, Cambridge, MA},\ \bibinfo
  {year} {1963})\BibitemShut {NoStop}%
\bibitem [{Note2()}]{Note2}%
  \BibitemOpen
  \bibinfo {note} {In Koster's notation\cite {Koster1963}, the irreducible
  representations (irreps) $\Gamma _i$ are usually labelled such that the
  low-symmetry irreps have small $i$ indices. The `+' symbol appearing as a
  subscript to the index in $\Gamma _i^+$ indicates that the irrep is symmetric
  with respect to the inversion.}\BibitemShut {Stop}%
\bibitem [{\citenamefont {Wybourne}(1965)}]{Wybourne1965}%
  \BibitemOpen
  \bibfield  {author} {\bibinfo {author} {\bibfnamefont {B.~G.}\ \bibnamefont
  {Wybourne}},\ }in\ \href@noop {} {\emph {\bibinfo {booktitle} {Spectroscopic
  properties of rare earths}}}\ (\bibinfo  {publisher} {Wiley, New York},\
  \bibinfo {year} {1965})\BibitemShut {NoStop}%
\bibitem [{\citenamefont {Boothroyd}(2015)}]{SPECTRE}%
  \BibitemOpen
  \bibfield  {author} {\bibinfo {author} {\bibfnamefont {A.~T.}\ \bibnamefont
  {Boothroyd}},\ }in\ \href@noop {} {\emph {\bibinfo {booktitle} {{SPECTRE, a
  program for calculating spectroscopic properties of rare earth ions in
  crystals}}}}\ (\bibinfo {year} {1990-2015})\BibitemShut {NoStop}%
\bibitem [{\citenamefont {Onoda}\ and\ \citenamefont
  {Tanaka}(2011)}]{Onoda2011}%
  \BibitemOpen
  \bibfield  {author} {\bibinfo {author} {\bibfnamefont {S.}~\bibnamefont
  {Onoda}}\ and\ \bibinfo {author} {\bibfnamefont {Y.}~\bibnamefont {Tanaka}},\
  }\href {\doibase 10.1103/PhysRevB.83.094411} {\bibfield  {journal} {\bibinfo
  {journal} {Phys. Rev. B}\ }\textbf {\bibinfo {volume} {83}},\ \bibinfo
  {pages} {094411} (\bibinfo {year} {2011})}\BibitemShut {NoStop}%
\bibitem [{\citenamefont {Bramwell}\ and\ \citenamefont
  {Harris}(1998)}]{Harris1998}%
  \BibitemOpen
  \bibfield  {author} {\bibinfo {author} {\bibfnamefont {S.~T.}\ \bibnamefont
  {Bramwell}}\ and\ \bibinfo {author} {\bibfnamefont {M.~J.}\ \bibnamefont
  {Harris}},\ }\href@noop {} {\bibfield  {journal} {\bibinfo  {journal} {J.
  Phys. Condens. Matter}\ }\textbf {\bibinfo {volume} {10}},\ \bibinfo {pages}
  {L215} (\bibinfo {year} {1998})}\BibitemShut {NoStop}%
\bibitem [{\citenamefont {Matsuhira}\ \emph
  {et~al.}(2002{\natexlab{a}})\citenamefont {Matsuhira}, \citenamefont {Hiroi},
  \citenamefont {Tayama}, \citenamefont {Takagi},\ and\ \citenamefont
  {Sakakibara}}]{Matsuhira2002}%
  \BibitemOpen
  \bibfield  {author} {\bibinfo {author} {\bibfnamefont {K.}~\bibnamefont
  {Matsuhira}}, \bibinfo {author} {\bibfnamefont {Z.}~\bibnamefont {Hiroi}},
  \bibinfo {author} {\bibfnamefont {T.}~\bibnamefont {Tayama}}, \bibinfo
  {author} {\bibfnamefont {S.}~\bibnamefont {Takagi}}, \ and\ \bibinfo {author}
  {\bibfnamefont {T.}~\bibnamefont {Sakakibara}},\ }\href@noop {} {\bibfield
  {journal} {\bibinfo  {journal} {J. Phys. Condens. Matter}\ }\textbf {\bibinfo
  {volume} {14}},\ \bibinfo {pages} {L559} (\bibinfo {year}
  {2002}{\natexlab{a}})}\BibitemShut {NoStop}%
\bibitem [{\citenamefont {Molavian}\ and\ \citenamefont
  {Gingras}(2009)}]{Molavian2009}%
  \BibitemOpen
  \bibfield  {author} {\bibinfo {author} {\bibfnamefont {H.~R.}\ \bibnamefont
  {Molavian}}\ and\ \bibinfo {author} {\bibfnamefont {M.~J.~P.}\ \bibnamefont
  {Gingras}},\ }\href@noop {} {\bibfield  {journal} {\bibinfo  {journal} {J.
  Phys. Condens. Matter}\ }\textbf {\bibinfo {volume} {21}},\ \bibinfo {pages}
  {172201} (\bibinfo {year} {2009})}\BibitemShut {NoStop}%
\bibitem [{\citenamefont {Machida}\ \emph {et~al.}(2010)\citenamefont
  {Machida}, \citenamefont {Nakatsuji}, \citenamefont {Onoda}, \citenamefont
  {Tayama},\ and\ \citenamefont {Sakakibara}}]{Machida2010}%
  \BibitemOpen
  \bibfield  {author} {\bibinfo {author} {\bibfnamefont {Y.}~\bibnamefont
  {Machida}}, \bibinfo {author} {\bibfnamefont {S.}~\bibnamefont {Nakatsuji}},
  \bibinfo {author} {\bibfnamefont {S.}~\bibnamefont {Onoda}}, \bibinfo
  {author} {\bibfnamefont {T.}~\bibnamefont {Tayama}}, \ and\ \bibinfo {author}
  {\bibfnamefont {T.}~\bibnamefont {Sakakibara}},\ }\href@noop {} {\bibfield
  {journal} {\bibinfo  {journal} {Nature}\ }\textbf {\bibinfo {volume} {463}},\
  \bibinfo {pages} {210} (\bibinfo {year} {2010})}\BibitemShut {NoStop}%
\bibitem [{\citenamefont {Matsuhira}\ \emph
  {et~al.}(2002{\natexlab{b}})\citenamefont {Matsuhira}, \citenamefont
  {Hinatsu}, \citenamefont {Tenya}, \citenamefont {Amitsuka},\ and\
  \citenamefont {Sakakibara}}]{Matsuhira2002b}%
  \BibitemOpen
  \bibfield  {author} {\bibinfo {author} {\bibfnamefont {K.}~\bibnamefont
  {Matsuhira}}, \bibinfo {author} {\bibfnamefont {Y.}~\bibnamefont {Hinatsu}},
  \bibinfo {author} {\bibfnamefont {K.}~\bibnamefont {Tenya}}, \bibinfo
  {author} {\bibfnamefont {H.}~\bibnamefont {Amitsuka}}, \ and\ \bibinfo
  {author} {\bibfnamefont {T.}~\bibnamefont {Sakakibara}},\ }\href {\doibase
  10.1143/JPSJ.71.1576} {\bibfield  {journal} {\bibinfo  {journal} {Journal of
  the Physical Society of Japan}\ }\textbf {\bibinfo {volume} {71}},\ \bibinfo
  {pages} {1576} (\bibinfo {year} {2002}{\natexlab{b}})}\BibitemShut {NoStop}%
\bibitem [{\citenamefont {Snyder}\ \emph {et~al.}(2004)\citenamefont {Snyder},
  \citenamefont {Ueland}, \citenamefont {Slusky}, \citenamefont {Karunadasa},
  \citenamefont {Cava},\ and\ \citenamefont {Schiffer}}]{Snyder2004}%
  \BibitemOpen
  \bibfield  {author} {\bibinfo {author} {\bibfnamefont {J.}~\bibnamefont
  {Snyder}}, \bibinfo {author} {\bibfnamefont {B.~G.}\ \bibnamefont {Ueland}},
  \bibinfo {author} {\bibfnamefont {J.~S.}\ \bibnamefont {Slusky}}, \bibinfo
  {author} {\bibfnamefont {H.}~\bibnamefont {Karunadasa}}, \bibinfo {author}
  {\bibfnamefont {R.~J.}\ \bibnamefont {Cava}}, \ and\ \bibinfo {author}
  {\bibfnamefont {P.}~\bibnamefont {Schiffer}},\ }\href {\doibase
  10.1103/PhysRevB.69.064414} {\bibfield  {journal} {\bibinfo  {journal} {Phys.
  Rev. B}\ }\textbf {\bibinfo {volume} {69}},\ \bibinfo {pages} {064414}
  (\bibinfo {year} {2004})}\BibitemShut {NoStop}%
\bibitem [{\citenamefont {Matsuhira}\ \emph {et~al.}(2011)\citenamefont
  {Matsuhira}, \citenamefont {Paulsen}, \citenamefont {Lhotel}, \citenamefont
  {Sekine}, \citenamefont {Hiroi},\ and\ \citenamefont
  {Takagi}}]{Matsuhira2011}%
  \BibitemOpen
  \bibfield  {author} {\bibinfo {author} {\bibfnamefont {K.}~\bibnamefont
  {Matsuhira}}, \bibinfo {author} {\bibfnamefont {C.}~\bibnamefont {Paulsen}},
  \bibinfo {author} {\bibfnamefont {E.}~\bibnamefont {Lhotel}}, \bibinfo
  {author} {\bibfnamefont {C.}~\bibnamefont {Sekine}}, \bibinfo {author}
  {\bibfnamefont {Z.}~\bibnamefont {Hiroi}}, \ and\ \bibinfo {author}
  {\bibfnamefont {S.}~\bibnamefont {Takagi}},\ }\href {\doibase
  10.1143/JPSJ.80.123711} {\bibfield  {journal} {\bibinfo  {journal} {Journal
  of the Physical Society of Japan}\ }\textbf {\bibinfo {volume} {80}},\
  \bibinfo {pages} {123711} (\bibinfo {year} {2011})}\BibitemShut {NoStop}%
\bibitem [{\citenamefont {Quilliam}\ \emph {et~al.}(2011)\citenamefont
  {Quilliam}, \citenamefont {Yaraskavitch}, \citenamefont {Dabkowska},
  \citenamefont {Gaulin},\ and\ \citenamefont {Kycia}}]{Quilliam2011}%
  \BibitemOpen
  \bibfield  {author} {\bibinfo {author} {\bibfnamefont {J.~A.}\ \bibnamefont
  {Quilliam}}, \bibinfo {author} {\bibfnamefont {L.~R.}\ \bibnamefont
  {Yaraskavitch}}, \bibinfo {author} {\bibfnamefont {H.~A.}\ \bibnamefont
  {Dabkowska}}, \bibinfo {author} {\bibfnamefont {B.~D.}\ \bibnamefont
  {Gaulin}}, \ and\ \bibinfo {author} {\bibfnamefont {J.~B.}\ \bibnamefont
  {Kycia}},\ }\href {\doibase 10.1103/PhysRevB.83.094424} {\bibfield  {journal}
  {\bibinfo  {journal} {Phys. Rev. B}\ }\textbf {\bibinfo {volume} {83}},\
  \bibinfo {pages} {094424} (\bibinfo {year} {2011})}\BibitemShut {NoStop}%
\bibitem [{\citenamefont {Jaubert}\ and\ \citenamefont
  {Holdsworth}(2011)}]{JaubertJPCM2011}%
  \BibitemOpen
  \bibfield  {author} {\bibinfo {author} {\bibfnamefont {L.~D.~C.}\
  \bibnamefont {Jaubert}}\ and\ \bibinfo {author} {\bibfnamefont {P.~C.~W.}\
  \bibnamefont {Holdsworth}},\ }\href@noop {} {\bibfield  {journal} {\bibinfo
  {journal} {Journal of Physics: Condensed Matter}\ }\textbf {\bibinfo {volume}
  {23}},\ \bibinfo {pages} {164222} (\bibinfo {year} {2011})}\BibitemShut
  {NoStop}%
\bibitem [{\citenamefont {Fennell}\ \emph {et~al.}(2012)\citenamefont
  {Fennell}, \citenamefont {Kenzelmann}, \citenamefont {Roessli}, \citenamefont
  {Haas},\ and\ \citenamefont {Cava}}]{Fennell2012}%
  \BibitemOpen
  \bibfield  {author} {\bibinfo {author} {\bibfnamefont {T.}~\bibnamefont
  {Fennell}}, \bibinfo {author} {\bibfnamefont {M.}~\bibnamefont {Kenzelmann}},
  \bibinfo {author} {\bibfnamefont {B.}~\bibnamefont {Roessli}}, \bibinfo
  {author} {\bibfnamefont {M.~K.}\ \bibnamefont {Haas}}, \ and\ \bibinfo
  {author} {\bibfnamefont {R.~J.}\ \bibnamefont {Cava}},\ }\href {\doibase
  10.1103/PhysRevLett.109.017201} {\bibfield  {journal} {\bibinfo  {journal}
  {Phys. Rev. Lett.}\ }\textbf {\bibinfo {volume} {109}},\ \bibinfo {pages}
  {017201} (\bibinfo {year} {2012})}\BibitemShut {NoStop}%
\bibitem [{\citenamefont {Lhotel}\ \emph {et~al.}(2012)\citenamefont {Lhotel},
  \citenamefont {Paulsen}, \citenamefont {de~R\'eotier}, \citenamefont
  {Yaouanc}, \citenamefont {Marin},\ and\ \citenamefont
  {Vanishri}}]{Lhotel2012}%
  \BibitemOpen
  \bibfield  {author} {\bibinfo {author} {\bibfnamefont {E.}~\bibnamefont
  {Lhotel}}, \bibinfo {author} {\bibfnamefont {C.}~\bibnamefont {Paulsen}},
  \bibinfo {author} {\bibfnamefont {P.~D.}\ \bibnamefont {de~R\'eotier}},
  \bibinfo {author} {\bibfnamefont {A.}~\bibnamefont {Yaouanc}}, \bibinfo
  {author} {\bibfnamefont {C.}~\bibnamefont {Marin}}, \ and\ \bibinfo {author}
  {\bibfnamefont {S.}~\bibnamefont {Vanishri}},\ }\href {\doibase
  10.1103/PhysRevB.86.020410} {\bibfield  {journal} {\bibinfo  {journal} {Phys.
  Rev. B}\ }\textbf {\bibinfo {volume} {86}},\ \bibinfo {pages} {020410}
  (\bibinfo {year} {2012})}\BibitemShut {NoStop}%
\bibitem [{\citenamefont {Takatsu}\ \emph {et~al.}(2012)\citenamefont
  {Takatsu}, \citenamefont {Kadowaki}, \citenamefont {Sato}, \citenamefont
  {Lynn}, \citenamefont {Tabata}, \citenamefont {Yamazaki},\ and\ \citenamefont
  {Matsuhira}}]{Takatsu2012}%
  \BibitemOpen
  \bibfield  {author} {\bibinfo {author} {\bibfnamefont {H.}~\bibnamefont
  {Takatsu}}, \bibinfo {author} {\bibfnamefont {H.}~\bibnamefont {Kadowaki}},
  \bibinfo {author} {\bibfnamefont {T.~J.}\ \bibnamefont {Sato}}, \bibinfo
  {author} {\bibfnamefont {J.~W.}\ \bibnamefont {Lynn}}, \bibinfo {author}
  {\bibfnamefont {Y.}~\bibnamefont {Tabata}}, \bibinfo {author} {\bibfnamefont
  {T.}~\bibnamefont {Yamazaki}}, \ and\ \bibinfo {author} {\bibfnamefont
  {K.}~\bibnamefont {Matsuhira}},\ }\href@noop {} {\bibfield  {journal}
  {\bibinfo  {journal} {Journal of Physics: Condensed Matter}\ }\textbf
  {\bibinfo {volume} {24}},\ \bibinfo {pages} {052201} (\bibinfo {year}
  {2012})}\BibitemShut {NoStop}%
\bibitem [{\citenamefont {Ehlers}\ \emph {et~al.}(2003)\citenamefont {Ehlers},
  \citenamefont {Cornelius}, \citenamefont {OrendÃ¡c}, \citenamefont
  {KajnakovÃ¡}, \citenamefont {Fennell}, \citenamefont {Bramwell},\ and\
  \citenamefont {Gardner}}]{Ehlers2002}%
  \BibitemOpen
  \bibfield  {author} {\bibinfo {author} {\bibfnamefont {G.}~\bibnamefont
  {Ehlers}}, \bibinfo {author} {\bibfnamefont {A.~L.}\ \bibnamefont
  {Cornelius}}, \bibinfo {author} {\bibfnamefont {M.}~\bibnamefont {OrendÃ¡c}},
  \bibinfo {author} {\bibfnamefont {M.}~\bibnamefont {KajnakovÃ¡}}, \bibinfo
  {author} {\bibfnamefont {T.}~\bibnamefont {Fennell}}, \bibinfo {author}
  {\bibfnamefont {S.~T.}\ \bibnamefont {Bramwell}}, \ and\ \bibinfo {author}
  {\bibfnamefont {J.~S.}\ \bibnamefont {Gardner}},\ }\href@noop {} {\bibfield
  {journal} {\bibinfo  {journal} {J. Phys. Condens. Matter}\ }\textbf {\bibinfo
  {volume} {15}},\ \bibinfo {pages} {L9} (\bibinfo {year} {2003})}\BibitemShut
  {NoStop}%
\bibitem [{\citenamefont {Ruminy}(2015)}]{Ruminy}%
  \BibitemOpen
  \bibfield  {author} {\bibinfo {author} {\bibfnamefont {M.}~\bibnamefont
  {Ruminy}},\ }\href@noop {} {\bibfield  {journal} {\bibinfo  {journal} {Thesis
  ETHZ}\ } (\bibinfo {year} {2015})}\BibitemShut {NoStop}%
\bibitem [{Note3()}]{Note3}%
  \BibitemOpen
  \bibinfo {note} {Heat capacity measurements providing a proof for a vanishing
  spin entropy at the temperatures where the persistent fluctuations are
  observed using neutron spectroscopy would be necessary to claim the quantum
  origin of the fluctuations.}\BibitemShut {Stop}%
\end{thebibliography}
\end{document}